\documentclass[letterpaper,preprint,11pt,aps,floatfix,amsmath,onecolumn]{revtex4-1}

\usepackage{amsmath}
\usepackage{epsfig}
\usepackage{array}
\usepackage{verbatim}
\usepackage{hyperref}
\usepackage{amssymb}
\usepackage{multirow}

\usepackage{color}
\usepackage{ulem}

\newcommand{\re}[1]{(\ref{#1})}

\newcommand{\beg}{\begin{equation}}

\newcommand{\en}{\end{equation}}

\newcommand {\dis}{\displaystyle}

\newcommand{\eps}{\varepsilon}
\newcommand{\lam}{\lambda}

\newcommand{\eref}[1]{Eq.~(\ref{#1})}
\newcommand{\esref}[1]{Eqs.~(\ref{#1})}

\newcommand{\up}{\uparrow}
\newcommand{\dn}{\downarrow} 

\renewcommand{\emph}{\textit}

\begin{document}

\title{Classification of parameter-dependent quantum integrable models% according to the number of independent integrals of motion%
, their parameterization, exact solution, and other properties}

\author{Haile K. Owusu and Emil A. Yuzbashyan}

\affiliation{Center for Materials Theory, Department of Physics and Astronomy,
Rutgers University, Piscataway, NJ 08854, USA}

\begin{abstract}
 
We study general quantum integrable Hamiltonians linear in a coupling constant and represented by finite $N\times N$ real symmetric matrices.  The restriction on the coupling dependence leads to a natural notion of nontrivial integrals of motion and classification of integrable families into {\it Types} according to the number of such integrals. A Type $M$ family in our definition is formed by $N-M$  nontrivial mutually commuting operators linear in the coupling.
Working from this definition alone, we parameterize Type $M$ operators, i.e. resolve the commutation relations, and  obtain an exact solution for their eigenvalues and eigenvectors. We  show that our parameterization covers all Type 1, 2, and 3 integrable models and discuss the extent to which it is complete for other types. We also present robust numerical observation on the number of energy level crossings in Type $M$ integrable systems and analyze the taxonomy of types in the 1d Hubbard model.
 
 \end{abstract}
 
\date{\today}
\maketitle

\tableofcontents

\newpage 
 
\section{\label{Intro} Introduction} 
 
Quantum integrability is usually defined as the existence of a sufficient number of nontrivial operators commuting with the Hamiltonian.  Such  operators  are interchangeably  referred to  as ``conserved currents", ``families of commuting Hamiltonians", integrals of motion, ``dynamical conservation laws", or simply ``dynamical symmetries", and are distinct from the ordinary space-time or internal space symmetries, e.g. momentum and spin conservation, particle-hole symmetry etc.  Quantum integrable systems are believed to have a range of  properties, which are sometimes proposed as alternative notions or tests of integrability -- an exact solution for their eigenstates, Poissonian energy level statistics, level crossings violating the  Wigner-von Neumann non-crossing rule in parameter dependent systems, and nondiffractive scattering,  see e.g. Ref.~\onlinecite{Caux} for a review.

There are two problems with the above definition. First, it has proved difficult to precisely formulate what constitutes a nontrivial commuting operator. For instance, any Hamiltonian commutes with projectors onto its eigenstates, so the mere statement that there exist commuting operators is not meaningful. In classical mechanics one requires functional independence of the integrals of motion\cite{arnold}. This criterion is not similarly useful in quantum mechanics where any two commuting matrices are functionally dependent unless there are degeneracies in the spectra of both of them\cite{Baum}.   
Second, in a system with a finite Hilbert space we clearly cannot have an infinite number of independent integrals, so we need to specify how many conserved currents are needed to say that the system is integrable. This is especially problematic in systems with no well-defined classical limit, e.g. the Hubbard model on a finite lattice\cite{hubbard,essler}, as, unlike the classical mechanics, there is no natural notion of the number of degrees of freedom. 
There is a similar lack of clarity in delineating the properties of quantum integrable systems. For example, there are poorly understood exceptions to the energy level statistics and level crossing tests\cite{Caux,owusu} as well as no convincing derivation of the exact solution and other properties {\it from} the notion of integrability. This is in sharp contrast to  the situation in classical mechanics where the Liouville-Arnold theorem\cite{arnold} demonstrates an exact solution by quadratures and quasi-periodic motion on invariant tori to be direct consequences of integrability.  

In the present paper we show that most of these problems can be resolved if one considers Hamiltonians and conserved currents that depend on a real parameter in a certain fixed way. This requirement alone leads to a well-defined notion of quantum integrability allowing us to classify and explicitly construct various integrable models, obtain their exact spectra, as well as derive other properties, directly from their integrability.
Examples of  such a real parameter, call it $u$,  in condensed matter integrable systems are: Coulomb interaction in the Hubbard model, magnetic field in Gaudin magnets with a boundary term\cite{sklyanin},  anisotropy in the $XXZ$ Heisenberg model, interaction strength in the  BCS model\cite{bcs} etc.  Dynamical symmetries depend on the same parameter in contrast  to ordinary symmetries which are $u$-independent.  Cases with no such parameter dependence, e.g. isotropic Heisenberg magnet and Gaudin model in zero field\cite{Gaudin}, often correspond to a more general integrable model at a specific value of the parameter.   In the Gaudin and BCS models  the Hamiltonian and all dynamical symmetries are linear in $u$\cite{sklyanin,dukelsky,cambiaggio}, while in the Hubbard and $XXZ$ models the Hamiltonian and the first conserved current are linear, while higher currents are higher order polynomials in $u$\cite{S,Lu,Gr,GM,zhou,Fu}.  

Note that in each of the above examples there is a subset of mutually commuting operators, including the Hamiltonian,  linear in the parameter. Let us therefore consider $N\times N$ Hermitian operators of the form $H(u)=T+uV$ and {\it require} the existence of dynamical symmetries linear in $u$. A simple, but important observation is that this restriction on $u$-dependence leads to a natural  notion of a nontrivial commuting partner. A typical (``non-integrable") $H(u)$ will commute only with a single (trivial) operator linear in $u$, namely with $\widetilde{H}(u)=(a+bu)I + c H(u)$, where $a, b$, and $c$ are real numbers and $I$ is an identity operator. 
The requirement that there exist a nontrivial commuting partner linear in $u$ imposes severe restrictions on $H(u)$. Indeed, as we will see below,  a real symmetric $H(u)$ for which such a partner exists is uniquely specified by a choice of no more than $4N$ real parameters, while for a generic $H(u)$ one needs $N(N+3)/2$ real parameters to fix the matrix elements of $T$ and $V$ \cite{count}.   Thus, Hermitian operators with fixed parameter-dependence separate into two distinct classes -- those that have nontrivial commuting partners and those that do not. 

Further, there is a natural classification of families of parameter-dependent commuting  operators according to the number $K$ of independent members they contain. Families with the maximum possible number $K=N$ (see below) of such operators we term Type 1, with one less than the maximum -- Type 2, etc. The main results of this paper are as follows. First, we  construct Type $M$ families of commuting operators $H^i(u)=T^i+u V^i$ for arbitrary $M$, i.e. we solve commutation relations $[H^i(u), H^j(u)]=0$ for matrix elements of $H^i(u)$. Second, we obtain exact eigenvalues and eigenstates of each $H^i(u)$. We thus derive the exact solution {\it from} the integrability. Third, we observe that the energy levels of $H^i(u)$ frequently cross and determine the number of level crossings as a function of $N$ and $M$.
 Type 1 (maximal) operators were previously constructed in Refs.~\onlinecite{shastry,owusu}. Here we analytically obtain {\it all} Type 2 and 3  and a  sub-class of Type $M$ real symmetric $N\times N$ operators for arbitrary $M>3$. In our construction a Type $M$ commuting family is parameterized by $2N+M-2$ real parameters, while, as we show, a generic Type $M$ involves $2N+2M-5$ parameters.
 
 There are typically $u$-independent symmetries common to the Hamiltonian and its dynamical symmetries, e.g. total momentum, particle number, $z$ component of the total spin, etc. It makes sense to ``factor" all these out, i.e. to go to blocks that correspond to a complete set of quantum numbers. Matrices we construct essentially represent  such ``irreducible" blocks of integrable Hamiltonians and their dynamical symmetries. As an example, we consider irreducible blocks for the Hubbard model with six sites, three spin-up and three spin-down electrons and determine their types.

A toy version of the above approach to integrability based on studying parameter-dependent commuting matrices  was discussed in Ref.~\onlinecite{emil} in the context of level crossings in the Hubbard model.   A major breakthrough came with Shastry's work \cite{shastry} where he constructed Type 1 matrices. The advantages of this approach are conceptual simplicity, clear meaning of independent integrals and freedom from specific details of particular integrable systems. A significant disadvantage is that the $H^i(u)$ correspond to isolated sectors/blocks of the Hamiltonian and dynamical symmetries and it is difficult to establish a correspondence between matrices  $H^i(u)$ and underlying operators whose blocks they represent.   Motivated by Shastry's work, we proposed a vector space definition (see Sect.~\ref{strategy}) and the above classification of integrable matrices into types, which allowed us to derive an explicit parameterization for Type 1 matrices, map them to Gaudin magnets and obtain an exact solution for their spectra \cite{owusu}.  Further, we demonstrated that $u$-dependent energy levels of Type 1 matrices have at least one and at most $(N-1)(N-2)/2$  crossings, while levels of integrable matrices of higher types do not have to cross, even though they frequently do. 

In what follows we first   lay out the general strategy for resolving the commutation relations in Sect.~\ref{strategy}. In Sect.~\ref{type1} we  briefly review the Type 1 construction of Refs.~\onlinecite{shastry,owusu}. Next, we obtain  all Type 2 matrices in Sect.~\ref{type2}. In Sect.~\ref{Type M} we construct integrable matrices of arbitrary Type $M$, which includes $M=1$ and 2 as particular cases.   Exact eigenvalues and eigenstates of Type $M$ are derived in Sect.~\ref{Exact}. Next, we present in Sect.~\ref{Gauge} `gauge' redundancies in the ansatz parameterization. In Sect.~\ref{numerics} numerical methods for generating random commuting matrices and determining their type, are presented and, thereby, the number of parameters necessary to specify a Type $M$ commuting family is determined. In this section we will also discuss   whether the construction of Sect.~\ref{Type M} generates all commuting families. Next, in Sect.~\ref{xings}, we look at the frequent level crossings that occur in Type $M$ matrices and relate their number to the properties of the discriminant of the characteristic polynomial of those matrices. We analyze the typology of various Hamiltonian blocks of the 1d Hubbard model in Sect.~\ref{Taxonomy}.  We conclude the main text with the summary and discussion of open questions. Finally, various proofs, details, and tangential results are relegated to  appendices \ref{typeMproof} through \ref{block}.

\section{\label{strategy}Resolving the commutation relations}

We start with primary definitions and  general observations. First, we define Type $M$ operators.
As outlined in the introduction, the task of identifying commuting operators linear in a real parameter $u$ in matrix representation reduces to finding a certain number, $K$, independent $N\times N$ Hermitian matrices $H^i(u)=T^i+u V^i$ such that 
\beg
[H^i(u),H^j(u)]=0\quad\mbox{for all $u$ and $i, j=0,\dots,K-1$}.
\label{comm}
\en
Equating terms at all orders of $u$, we have
\beg
[T^i, V^j]=[T^j,V^i],\quad [T^i, T^j]=[V^i, V^j]=0.
\label{comm1}
\en
As discussed above, we also require that there be no $u$-independent symmetry common to all $H^i(u)$, i.e. there is no $\Omega \neq aI$ such that $ [\Omega,H^i(u)]=0$ for all $u$ and $i$, where $I$ is an $N\times N$  identity matrix. When such symmetry is present, $H^i(u)$ acquire  block-diagonal structure and the problem reduces to that of smaller matrices. Note also that every commuting family contains the trivial member 
\beg
H^0(u)=(a+bu)I,
\label{h0}
\en 
where $a$ and $b$ are arbitrary real numbers. 

In light of the restriction on the $u$-dependence, being independent of $H^i(u)$ with $i=0,\dots, K-1$ simply means linear independence, i.e. $\sum_i a_i H^i(u)$ with real $a_i$ is zero if and only if all $a_i=0$. Note that due to the absence of $u$-independent symmetries linear independence of $H^i(u)$ is equivalent to that of $V^i$ or $T^i$ separately. 
Further, $K$ is defined as the maximum number of independent commuting matrices in a given family. In other words, any $H(u)=T+u V$ that commutes with all $H^i(u)$  is linearly dependent on those same $H^i(u)$, i.e.
\beg
H(u)=\sum_{i=0}^{K-1} a_i H^i(u).
\label{vecspace}
\en
We see that a  commuting family is a $K$-dimensional vector space ${\cal V}_K$ where $H^i(u)$ serve as basic vectors. There is a certain freedom in choosing a basis in the vector space ${\cal V}_K$, which will be exploited in Sects.~\ref{type1} and \ref{type2}.

It turns out that a typical  matrix of the form $T+uV$ has no nontrivial independent commuting partners linear in $u$ \cite{owusu}; it commutes only with $H^0(u)=(a+b u)I$ and itself.  Commutation relations \eref{comm} impose  strong constraints on  the matrix elements of {\it each} $H^i(u)$   such that  the probability of finding    randomly generated matrices $T+uV$   with nontrivial commuting  partners is zero. For example, in the $3\times 3$ case matrix elements of $T$ and $V$ must meet a single algebraic constraint for such a partner to exist \cite{emil} and the number of constraints increases with the size of the matrix. In our experience, finding an operator with even one nontrivial commuting partner is quite difficult. A  brute force approach to generating all such operators numerically (see Sect.~\ref{numerics}) already becomes computationally prohibitive for $N\ge 6$.

Therefore, the minimum (and  typical) number of independent  $H^i(u)$ in the family is $K=2$. The maximum possible number can be shown \cite{owusu} to be $K=N$, i.e. $K$ ranges from 2 to $N$. Following  Ref.~\onlinecite{owusu} we term the commuting families with the maximum number $N$ of $H^i(u)$ -- Type 1, those with $N-1$ commuting operators -- Type 2, etc. up to Type $N-1$. An arbitrary Type $M$   family is one with $K=N-M+1$ independent  commuting $N\times N$ matrices $H^i=T^i+uV^i$, which consist of the trivial independent element 
$H^0(u)=(a+b u)I$ and $K-1$ nontrivial ones.

Thus, our task is to identify nontrivial solutions of commutation relations \re{comm1}. An important observation, which considerably simplifies this task, is that we have freedom to choose bases in two distinct linear spaces without loss of generality. First, we can choose a basis in the ``target'' space -- the Hilbert space ${\cal H}_N$ on which Hermitian operators $T^i$ and $V^i$ act. Indeed, commutation relations \re{comm1} are invariant with respect to transformations $(T^i, V^i) \to (U^\dagger T^i U, U^\dagger V^i U)$, where $U$ is any $u$-independent unitary operator. Second, we have freedom to choose a basis in the vector space ${\cal V}_K$ of commuting operators, i.e. to make linear transformations $H^i(u) \to \sum_j a^i_j H^j(u)$ with any real, $u$-independent and non-degenerate  matrix $(a^i_j)$.

It is convenient to choose the basis in the target space to be the common eigenbasis of the  mutually commuting matrices $V^i$. In this basis, $V^i$ are diagonal matrices whose nonzero entries we will denote $d^i_k$, i.e.
\beg
\mbox{diagonal of $V^i$}= \left(d^i_1, d^i_2,\dots, d^i_N\right).
\label{vi}
\en
The commutation relation $[T^i, V^j]=[T^j,V^i]$ in this basis reads $( d^i_k-d^i_m)T^j_{km}=( d^j_k-d^j_m)T^i_{km}$. This implies that an auxiliary antisymmetric matrix  $S$ whose matrix elements are
\beg
S_{km}\equiv \dfrac{T^i_{km}}{d^i_k-d^i_m}=\dfrac{T^j_{km}}{d^j_k-d^j_m}
\label{Sdef}
\en
 does not depend on $i$, i.e. is the same for all members of the family.  Moreover, it follows from Eq.~\re{Sdef} that
\beg
T^i = W^i+ \left[ V^i,S \right]
\label{T}
\en
where $W^i$ is a diagonal matrix (the diagonal of $T^i$)\cite{shastry}. Its nonzero elements we denote $f^i_k$,
\beg
\mbox{diagonal of $T^i$}= \left(f^i_1, f^i_2,\dots, f^i_N\right).
\label{wi}
\en
Now the commutation relations $[V^i,V^j]=0$ and $[T^i, V^j]=[T^j,V^i]$ are automatically satisfied and \esref{comm1} reduce to a single equation $[T^i, T^j]=0$ or, equivalently, to 
\beg
\left[ [V^i,S], [V^j,S] \right]+ \left[[V^i,S],W^j \right]-\left[[V^j,S],W^i \right]=0.
\label{comm2}
\en
We have, for fixed $i$ and $j$, a single  matrix ``master" equation involving $4N+N(N-1)/2$ variables -- $4N$ corresponding to $V^i,V^j,W^i,W^j$ and $N(N-1)/2$ corresponding to $S$ -- and $N(N-1)/2$ constraints. In addition to this master equation, we will find that the quantity
\beg
\nu_{ijkl} \equiv S_{ij} S_{jk} S_{kl} S_{li}+S_{ik} S_{kl} S_{lj} S_{ji}+S_{il} S_{lj} S_{jk} S_{ki},
\label{nu1234}
\en
is an important one for determining matrix type. We will see that for Type 1, $\nu_{ijkl}=0$, and for Type 2, $\nu_{ijkl}=G x_i x_j x_k x_l$ where $G$ is some real parameter and $x_j$, $j=\{1,\dots,N \}$ are related to the matrix elements of $V^j$.

Next, let us discuss the choice of basis in the vector space ${\cal V}_K$ of $K$ commuting operators. Note from Eq.~\re{T} that a linear transformation  $H^i(u) \to \sum_j a^i_j H^j(u)$ translates into the same transformation on diagonal matrices $V^i$ and $W^i$ and does not affect the matrix $S$, i.e. $S$ is independent of the choice of basis in ${\cal V}_K$ and in this sense is a universal characteristic of the commuting family. Moreover, we will see that for Types 1 and 2 the master Eq.~\re{comm2} reduces to a single equation on matrix elements of $S$. Once $S$ is fully parameterized, $V^i$ and $W^i$ can be determined from Eq.~\re{comm2}. For Type $M>2$ families, we do not attempt to resolve \eref{comm2} generally, but use the Type 1 and 2 resolutions to formulate a working ansatz that generates such families. 

As mentioned above, linear independence of $H^i(u)$ is equivalent to that of $V^i$. Given Eq.~(\ref{vi}), this means that $K=N-M+1$ vectors $\vec d^i$  are linearly independent. Then, via a linear transformation $V^i \to \sum a^i_j V^j$ we can go to a \emph{canonical} basis in the operator space ${\cal V}_K$ such that $K-1=N-M$ of diagonal matrix elements $d^i_k$ are zero for every $i$. This is the basis we will use for Type 1 and Type 2 families in Sects.~\ref{type1} and \ref{type2}, respectively. 

All Hermitian Type 1 matrices were explicitly parameterized in Ref.~\onlinecite{owusu}. In constructing higher types here
we will  restrict our analysis to real symmetric as opposed to general Hermitian matrices  for simplicity. Specifically, we will choose the defining parameters of Type $M$ commuting families to be real and such that the resulting matrices are real symmetric. We note however that all other properties of Type $M$ operators we construct in this paper -- i.e. commutation relations \re{comm1} and exact solution for the spectra of $H^i(u)$ (see below) -- persist even when the parameters are unrestricted complex numbers. 

\section{\label{type1} Review of Type 1  matrices}

Here we review the main results for Type 1 \cite{owusu,shastry} -- families of $N \times N$ commuting operators, linear in a parameter $u$ that have a maximum number ($N$) of linearly independent members, which were also termed   ``maximal operators" in Ref.~\onlinecite{owusu}. As discussed in the previous section, we can go from a general operator basis  to a convenient canonical  basis by taking linear combinations of $H^i(u)=T^i+uV^i$. Since there are $N$ linearly independent diagonal matrices $V^i$ in the case of Type 1, we can choose the basis so that all $d^i_k$ in Eq.~\re{vi} are zero except one, i.e. $V^i_{kk}\equiv d^i_k=\delta_{ik}$ for all $i$.

Consider \eref{comm2} in the Type 1 canonical basis. It turns out that the left hand side of this equation is an explicitly mostly empty matrix. Evaluating its nonzero matrix elements, we obtain after some algebra
\beg
f^i_j- f^i_k=-\dfrac{S_{ij} S_{ik}}{S_{jk}},
\label{f's1}
\en
where $i, j, k$ are distinct and $f^i_k$ are the diagonal matrix elements of $T^i$, see Eq.~\re{wi}.
Consistency requires that a triangular sum of such relations itself vanish, i.e. $(f^i_j-f^i_k)+(f^i_k-f^i_l)+ (f^i_l-f^i_j)=0$, from which follows
\beg
\nu_{ijkl}=0,
\label{nu1}
\en 
where $\nu_{ijkl}$ is defined in \eref{nu1234}.
This is the main equation defining Type 1 matrices\cite{shastry}. The key insight in Ref.~\onlinecite{owusu} is the understanding that \eref{nu1} \emph{necessarily} implies the parameterization
\beg
S_{jk}=\dfrac{\gamma_j \gamma_k}{\varepsilon_j-\varepsilon_k},
\label{S1}
\en
with unrestricted real parameters $\gamma_j$ and $\varepsilon_j$ characteristic of the commuting family (see also Ref.~\onlinecite{shastry1} for an elegant alternative derivation of \eref{S1} with the help of Pl\"ucker relations).

Once the matrix $S$ is known, it is straightforward to calculate $f^i_k$ and matrices $T^i$ with the help of Eqs.~(\ref{comm2}) and (\ref{f's1}).  A general Type 1 matrix $H(u)=\sum_{i=1}^N d_i H^i(u)$ with arbitrary real $d_i$ takes the form
\begin{equation}
\begin{array}{l}
\dis \left[H \left(u\right)\right]_{mn}= \gamma_{m}\gamma_{n}
\left(\frac{d_{m} -d_{n} }{\varepsilon_{m}-\varepsilon_{n}}\right),\quad m\ne n,\\
\\
\dis\left[H \left(u\right)\right]_{mm}=u\, d_{m} -\sum_{j\neq
m}\gamma_{j}^{2}
\left(\frac{d_{m} -d_{j} }{\varepsilon_{m}-\varepsilon_{j}}\right).\\
\end{array}
\label{Hmel}
\end{equation}
  Eq.~\re{Hmel} provides a complete explicit parameterization of all Type 1 matrices. Note also that the parameterization \re{Hmel} is independent of the choice of basis in the operator space ${\cal V}_K$.

In Ref.~\onlinecite{owusu}  we further show that Type 1 operators map onto a sector of Gaudin magnets in the presence of external magnetic field  and as a consequence of this correspondence, we use the well known exact solution of the Gaudin model\cite{Gaudin,sklyanin,dukelsky} to furnish an exact solution for Type 1. The components of eigenvector $\vec{v}_m (u)$ of Type 1 matrix $H(u)$ are
\beg
  \left[ \vec{v}_m (u) \right]_j= \dfrac{\gamma_j}{\lambda_m-\varepsilon_j},
\label{eigenvector1}
\en
corresponding to eigenvalue
\beg
E_m(u)=\sum_{k=1}^N{\dfrac{d_k\gamma_k^2}{\lambda_m-\varepsilon_k}},
\en
where the $\lambda_i$, $i=1, \dots, N$ are subject to a single algebraic equation
\beg
 \sum_{j=1}^N{\dfrac{\gamma_j^2}{\lambda_m-\varepsilon_j}}=u.
 \label{constraint1}
\en
An intriguing consequence of this exact solution is that a Type 1 $H(u)$ typically multiply violates the Wigner-von Neumann non-crossing rule\cite{Hund,Neumann,Teller,Landau,Longuet-Higgins,Naqvi,Kestner} and, moreover, it is \emph{necessary} that it do so at least once\cite{owusu}.

\section{\label{type2}Complete parameterization of Type 2 matrices}

In this section we construct all Type 2 families, i.e. linear vector spaces formed by $N-1$ mutually commuting, independent matrices $H^i(u)=T^i+u V^i$. The derivation is similar to that of Type 1 in the previous section, but somewhat more involved. We first choose a convenient canonical basis in the operator space ${\cal V}_K$ as discussed in the end of Sect.~\ref{strategy} and evaluate matrix elements of \eref{comm2} in this basis. This results in an equation constraining the matrix $S_{kl}$ defining Type 2 matrices, which we then solve. The final result -- parameterization of Type 2 operators -- is given by \eref{Hmel2}.

There are $N-1$ linearly independent diagonal matrices $V^i$ for any Type 2 family. Therefore, by taking linear combinations of $H^i(u)$, we can go to a basis in the operator space where only two of the $d^i_k$ in Eq.~\re{vi} are nonzero. It is convenient to choose and parameterize these nonzero matrix elements as $d^i_j=1/x_j$ for some fixed $j$ and $d^i_i=-1/x_i$ for $i=1, 2, \dots, j-1, j+1,\dots, N$. We also impose the following constraint on parameters $x_k$:
\beg
\sum_{k=1}^N {x_k}=0,
\label{x's}
\en
so that the identity can be formed by linear combinations of the $H^i(u)$, thereby ensuring that it is an element of the commuting family.

\eref{comm2} in this basis yields
\beg
f^{i}_k-f^{i}_l=- \dfrac{S_{jk} S_{jl}}{x_j S_{kl}} + \dfrac{S_{ik} S_{il}}{x_i S_{kl}},
\label{diagII}
\en
for $k,l \neq \lbrace i,j \rbrace$. Similar to the analogue \eref{f's1}, consistency requires that a triangular sum of such terms vanish, i.e. $(f^i_{k}-f^i_{l})+(f^i_{l}-f^i_{m})+(f^i_{m}-f^i_{k})=0$, from which follows that $x_i \nu_{jklm}=x_j \nu_{iklm}$. Without loss of generality we can define a quantity $G_{ijkl}$ such that $\nu_{ijkl}=G_{ijkl}x_i x_j x_k x_l$, which implies that $G_{jklm}=G_{iklm}$. This relation is general and must be true for all distinct $i,j,k,l,m$, and therefore consistently permuting indices yields $G_{ijkl}=\mbox{const}=G$, where $G$ is  some nonzero real parameter. We therefore obtain
\beg
\nu_{ijkl} =G x_i x_j x_k x_l,
\label{S's}
\en 
 We will show that \eref{S's} can be solved generically. Let us simplify  by dividing out the $x$'s, i.e. let $S_{ij} \equiv \widetilde{S}_{ij}  \sqrt{x_i}\sqrt{ x_j}$, from which it follows that 
\beg
\widetilde{\nu}_{ijkl}\equiv \widetilde{S}_{ij} \widetilde{S}_{jk} \widetilde{S}_{kl} \widetilde{S}_{li}+\widetilde{S}_{ik} \widetilde{S}_{kl} \widetilde{S}_{lj} \widetilde{S}_{ji}+\widetilde{S}_{il} \widetilde{S}_{lj} \widetilde{S}_{jk} \widetilde{S}_{ki}=G.
\label{nu}
\en
 
We are looking for a parameterization of $\widetilde{S}_{ij}$ that resolves \eref{nu}, ideally in a manner that treats both of its indices similarly. Our strategy for pursuing this parameterization is to find another equation nontrivially related to \eref{nu} and use that to eliminate some of the $S$'s. More specifically, \eref{nu} involves four indices whereas the $S$'s are two index objects -- implying that \eref{nu} overdetermines the $S$'s. We are looking for an equation involving fewer indices from which to derive a parameterization. 

The $\widetilde{S}_{ij}$, like the $S_{ij}$, are antisymmetric and, as it turns out, this property is sufficient to identically satisfy the equation:
\[
\widetilde{\nu}_{mjkl} \widetilde{S}_{ij} \widetilde{S}_{ik} \widetilde{S}_{il}-\widetilde{\nu}_{mikl} \widetilde{S}_{ij} \widetilde{S}_{jk} \widetilde{S}_{jl}+\widetilde{\nu}_{mijl} \widetilde{S}_{ik} \widetilde{S}_{jk} \widetilde{S}_{kl}-\widetilde{\nu}_{mijk} \widetilde{S}_{il} \widetilde{S}_{jl} \widetilde{S}_{kl}=0.
\]
Given \eref{nu}, we can remove an overall common factor $G$ such that
\beg
\widetilde{S}_{ij} \widetilde{S}_{ik} \widetilde{S}_{il}- \widetilde{S}_{ij} \widetilde{S}_{jk} \widetilde{S}_{jl}+ \widetilde{S}_{ik} \widetilde{S}_{jk} \widetilde{S}_{kl}- \widetilde{S}_{il} \widetilde{S}_{jl} \widetilde{S}_{kl}=0.
\label{rho n nu}
\en
\eref{rho n nu} is now a nontrivial equation constraining the $\widetilde{S}_{ij}$ beyond simple antisymmetry.  We can use \eref{nu} and \eref{rho n nu} to eliminate $\widetilde{S}_{kl}$. This is the resulting equation 
\begin{equation}
\begin{split}
\rho_{ij} &\equiv \dfrac{\left( \widetilde{S}_{ij} \widetilde{S}_{ik} \right)^2+\left( \widetilde{S}_{ij} \widetilde{S}_{jk} \right)^2-\left( \widetilde{S}_{ik} \widetilde{S}_{jk} \right)^2-G}{2\widetilde{S}_{ik} \widetilde{S}_{jk}}+\widetilde{S}_{ij}^2=
\\
&\phantom{\equiv a}\dfrac{\left( \widetilde{S}_{ij} \widetilde{S}_{il} \right)^2+\left( \widetilde{S}_{ij} \widetilde{S}_{jl} \right)^2-\left( \widetilde{S}_{il} \widetilde{S}_{jl} \right)^2-G}{2\widetilde{S}_{il} \widetilde{S}_{jl}}+\widetilde{S}_{ij}^2,
\end{split}
\label{rhos}
\end{equation}
where we have chosen to separate terms involving index $k$ and those involving index $l$. Note that both sides of the equation are identical but for the different indices, true for all $k,l \neq \{ i,j \}$, and it is for this reason that we can define the object $\rho_{ij}$ as a two-index quantity though it is defined through quantities involving three indices.   From \eref{rhos} we obtain
\beg
\widetilde{S}_{ij}^2 \left( \widetilde{S}_{ik}+\widetilde{S}_{jk} \right)^2-\left( \widetilde{S}_{ik}\widetilde{S}_{jk} +\rho_{ij} \right)^2=G-\rho_{ij}^2,
\label{Type II master}
\en
for all $k \neq \{i,j\}$. The key result of our strategy  is that \eref{Type II master} implies that
\begin{equation}
\begin{array}{l}
\dis \widetilde{S}_{ik}+\widetilde{S}_{jk}=\dfrac{\sqrt{G-\rho_{ij}^2}}{\widetilde{S}_{ij}} \cosh{\chi_k},
\\
\\
\dis \widetilde{S}_{ik}\widetilde{S}_{jk}=\sqrt{G-\rho_{ij}^2} \sinh{\chi_k}-\rho_{ij},
\\
\\
\end{array}
\label{solved}
\end{equation}
for some parameters $\chi_k$. 

Again, what we have done here is use the identically satisfied \eref{rho n nu} to reduce a four-index equation \re{nu} to a number of three-index equations involving the index-less parameter $G$ and the two-index object $\rho_{ij}$. Actually, at this point, there is a simple algorithm to completely determine all $S$'s. We can fix indices $i$ and $j$ and choose values for $G$, $\rho_{ij}$, $\widetilde{S}_{ij}$ and these new parameters $\chi_k$, $k \neq \{i,j\}$. Given these chosen values, the $\widetilde{S}_{ik}$ and $\widetilde{S}_{jk}$ can be determined using \eref{solved}. Once we have values for $\widetilde{S}_{ik}$ and $\widetilde{S}_{jk}$, substitution into \eref{rho n nu} allows us to completely determine $\widetilde{S}_{kl}$, $k,l \neq \{i,j\}$, i.e.
\beg
\widetilde{S}_{kl}=-\widetilde{S}_{ij}\dfrac{\widetilde{S}_{ik}\widetilde{S}_{il}-\widetilde{S}_{jk}\widetilde{S}_{jl}}{\widetilde{S}_{ik}\widetilde{S}_{jk}-\widetilde{S}_{il}\widetilde{S}_{jl}}=-\dfrac{\widetilde{ \Gamma}_l \cosh{\chi_k}+\widetilde{ \Gamma}_k \cosh{\chi_l}}{\sinh{\chi_k}-\sinh{\chi_l}},
\label{retroS}
\en
where 
\[
\widetilde{\Gamma}_m \equiv \delta_m \sqrt{\frac{G-\rho_{ij}^2}{\widetilde{S}_{ij}^2} \cosh{\chi_m}^2-4\sqrt{G-\rho_{ij}^2}\sinh{\chi_m}+4\rho_{ij}},
\]
and $\delta_m=\pm 1$ accounts for an inherent ambiguity in sign.

The above algorithm explicitly favors two indices over the others, i.e. the fixed $i$ and $j$ above, and we would like to treat all indices in a unified way. Let us make the following change of variables: $\sinh{\chi_m} =(q\varepsilon_m+r)/(s\varepsilon_m+t)$ for $q,r,s,t$ such that $q t-r s=1$ \cite{normalization}, and $m \neq \{i,j\}$. By substituting this into \eref{retroS} and removing an overall scale (again without loss of generality), we find that
\beg
\widetilde{S}_{kl}=\dfrac{1}{2} \sqrt{\left( \lambda_1-\varepsilon_k \right) \left( \lambda_2-\varepsilon_k \right) \left( \lambda_1-\varepsilon_l \right) \left( \lambda_2-\varepsilon_l \right)} \;\dfrac{\Gamma_k+\Gamma_l}{\varepsilon_k-\varepsilon_l},
\label{tilde form}
\en
given
\beg
\Gamma_m \equiv \Gamma(\varepsilon_m)=\delta_m \sqrt{1+\frac{P_1}{\lambda_1-\varepsilon_m}+\frac{P_2}{\lambda_2-\varepsilon_m}},
\label{GamGam}
\en
where the dependence of $\lambda_1, \lambda_2, P_1,P_2$ on parameters $q,r,s,t,G,\rho_{ij},\widetilde{S}_{ij}$ can be determined with a bit of algebra. We are not particularly interested in this functional dependence, however, because \eref{GamGam} generally satisfies \eref{nu}, i.e. if we choose parameters $\lambda_1, \lambda_2, P_1,P_2$ and $\varepsilon_m$ for $1 \leq m \leq N$, \emph{all} $\widetilde{S}_{kl}$ given by \eref{tilde form}, $k,l=\{1 \dots N\}$, satisfy \eref{nu} where in these preferred parameters $G=-P_1 P_2 (\lambda_1-\lambda_2)^2/16$.

  To define the Type 2 $S_{ij} \equiv \widetilde{S}_{ij}  \sqrt{x_i}\sqrt{ x_j}$  fully, we can simply choose $N-1$ of the $x_j$ parameters and determine the remaining one so that \eref{x's} is satisfied. We can, however, write $x_j$ in a more transparent form. Let us   introduce new real parameters $\gamma_j$ such that
\beg
x_j = \dfrac{\gamma_j^2}{\left( \lambda_1-\varepsilon_j \right) \left( \lambda_2-\varepsilon_j \right)}.
\en
 By partial fraction decomposition 
\beg
\dfrac{\gamma_j^2}{\left( \lambda_1-\varepsilon_j \right) \left( \lambda_2-\varepsilon_j \right)}=-\dfrac{1}{\lambda_1-\lambda_2} \left( \dfrac{\gamma_j^2}{\lambda_1-\varepsilon_j}-\dfrac{\gamma_j^2}{\lambda_2-\varepsilon_j}  \right),
\label{Gaudin sep}
\en
we see that we can impose \eref{x's} by requiring that $\lambda_1$ and $\lambda_2$ be among the $N$ solutions to 
\beg
\sum_{j=1}^{N}\dfrac{\gamma_{j}^2}{\lambda_i-\varepsilon_j}=B,
\label{Gaudin1}
\en
where $B$ is an arbitrary real number and $\varepsilon_m$ are understood to be distinct.
Given \eref{tilde form} we find that Type 2 $S_{kl}$ take the simple form
\beg
S_{kl}=\dfrac{1}{2} \dfrac{\gamma_k \gamma_l}{\varepsilon_k-\varepsilon_l}\left( \Gamma_k+\Gamma_l \right).
\label{full form}
\en

Fully parameterizing Type 2 $S_{kl}$ is the lions' share of the  Type 2 parameterization effort and with a bit of algebraic manipulation we determine the $f^{i}_k$ , for all $k$ using Eqs.~(\ref{comm2}) and (\ref{diagII}), which are linear in $f^{i}_k$.  Linear combinations of the resulting $H^{i}(u)$ yield a general Type 2 matrix
\begin{equation}
\begin{array}{l}
\dis \left[H \left(u\right)\right]_{mn}=
\gamma_{m}\gamma_{n} \left(\frac{d_{m} -d_{n} }{\varepsilon_{m}-\varepsilon_{n}}\right) \dfrac{\Gamma_m +\Gamma_n}{2},\quad m\ne n,\\
\\
\dis\left[H \left(u\right)\right]_{mm}=u\, d_{m} -\sum_{j\neq m} {\gamma_{j}^{2}
\left(\frac{d_{m} -d_{j} }{\varepsilon_{m}-\varepsilon_{j}}\right) \dfrac{1}{2} \dfrac{\left( \Gamma_m+\Gamma_j\right)\left( \Gamma_j+1\right)}{\Gamma_m+1}},\,\
\end{array}
\label{Hmel2}
\end{equation}
where 
\beg
d_m=g_0+\sum_{j=1}^{N-2}\frac{g_j}{\lambda_{j+2}-\varepsilon_m}
\label{dm2}
\en
and $g_j$ are arbitrary real numbers.  

 \eref{Hmel2} together with auxiliary   \esref{GamGam}, (\ref{Gaudin1}) and \re{dm2} provide a complete parameterization of all Type 2 commuting families. $2N+3$ independent real parameters $B$, $P_1$, $P_2$, $\gamma_i$ and $\varepsilon_i$ specify the commuting family; $N-1$ real numbers $g_i$ further specify a particular matrix within the family. The choice of all these parameters is unconstrained, except simple explicit restrictions on $P_1$ and $P_2$ to ensure that \eref{GamGam} produces real $\Gamma_m$, which we discuss in detail in the next section for general Type $M$.
 
 The choice of parameters is not unique. For example, a uniform shift $\varepsilon_i \to \varepsilon_i + c$ yields the same family. The entire group of ``gauge" transformations on the parameters that leave Type 2 commuting families invariant is explored in Sect.~\ref{2gauge}. Interestingly, it turns out that in the case of Type 2 there is an additional gauge freedom not present for other types, which allows an alternative, particularly transparent parameterization, see \eref{ellS}.

\section{\label{Type M}Type $M$ - the ansatz parameterization}

Explicitly constructing the most general Type $M \geq 3$ commuting families is difficult if at all possible (see the discussion  Sect.~\ref{discussion} for more on this). However, it turns out that the Type 2 formulas of the previous section can be generalized to obtain some Type $M$  families with arbitrary $M$. The idea is to use the same Eqs.~\re{full form} and  \re{Hmel2} to generate the matrices, but to extend 
\eref{GamGam} by including more poles in its radicand. Specifically, we take   
\beg
\Gamma_m \equiv \Gamma(\eps_m),\qquad \Gamma(\sigma)= \delta(\sigma) \sqrt{1+\sum_{j=1}^{M} \frac{P_j}{\lambda_{j}-\sigma}}, 
\label{gams}
\en
\beg
d_m=g_0+\sum_{j=1}^{N-M}{\frac{g_{j}}{\lambda_{j+M}-\varepsilon_m}},
\label{dmM}
\en
where $\delta(\sigma)=\pm1$, $P_j$  are real numbers and $\lambda_m$, $m=\{ 1,\dots N\}$ are solutions of \eref{Gaudin1}. 

In Appendix~\ref{typeMproof} we prove that \eref{Hmel2} with $\Gamma_m$ and $d_m$ given by Eqs.~\re{gams} and \re{dmM}, respectively, indeed produce Type $M$ matrices as defined in Sec.~\ref{strategy}, i.e. families of $K=N-M+1$ linearly independent mutually commuting matrices with no $u$-independent symmetry. Eqs.~\re{Hmel2} and \re{gams} contain $2N+M+1$  parameters characterizing the commuting family -- $2N$  $\gamma_i$'s and $\varepsilon_i$'s, $M$ $P_i$'s and parameter $B$. As discussed in detail in Sect.~\ref{Gauge}, there are certain gauge transformations, such as a uniform scaling of $\gamma_i$ and $\varepsilon_i$ or a uniform shift of $\varepsilon_i$,  that leave the commuting family invariant.  This allows to fix three  degrees of freedom meaning that the number of parameters needed to uniquely specify a Type $M$ commuting family produced by our ansatz is $2N+M-2$. In Sects.~\ref{numerics} and \ref{discussion} we argue based on numerical evidence and other considerations that a {\it general} Type $M \geq 3$ family is uniquely specified by $2N+2M-5$ parameters. This implies that our construction misses $M-3$ parameters, i.e. it does not yield all commuting families for types $M>3$.

 A natural way to choose a basis for the ansatz Type $M$ commuting family   is to define the $K-1=N-M$ nontrivial $H^i(u)$ such that    $g_j=\delta_{ij}$ in \eref{dmM} for  $1\le i\le N-M$. For $i=0$, we take $H^0(u)$ to be proportional to the identity matrix, as given in \eref{h0}.  In other words,
 \beg
\mbox{$H^i(u)=T^i+uV^i$ is given by Eqs.~\re{Hmel2} and \re{gams} with  $\displaystyle d_k\to d_k^i=\frac{1}{\lambda_{i+M}-\varepsilon_k}$}.
 \label{him}
 \en
 for $i=1,\dots,N-M$. In particular,
  $V^i$ is a diagonal matrix and
 \beg
\mbox{diagonal of $V^i$}= (d_1^i, d_2^i,\dots,d_N^i)=
 \left(\frac{1}{\lambda_{i+M}-\eps_1}, \frac{1}{\lambda_{i+M}-\eps_2},\dots,\frac{1}{\lambda_{i+M}-\eps_N}\right).
 \label{vibasis}
 \en
A general member of the commuting family is 
\beg
H(u)=\sum_{i=0}^{N-M} g_i H^i(u),
\label{genedfin}
\en
consistent with Eqs.~\re{Hmel2} and \re{dmM}, where $g_i$ are arbitrary real coefficients.

Especially interesting in the structure of the ansatz is the form of \eref{gams}. The presence of these $\Gamma_j$'s in \eref{Hmel2} is what essentially distinguishes Type $M$, $M>1$, from the Type 1 operators reviewed in Sec.~\ref{type1}, and therefore what allows for the systematic decrease in size of the commuting family, i.e. these $\Gamma$'s see us go from maximal operator $S_{kl}=\gamma_k \gamma_l/(\varepsilon_k-\varepsilon_l)$ to 
\beg
S_{kl}=\dfrac{1}{2} \dfrac{\gamma_k \gamma_l}{\varepsilon_k-\varepsilon_l}\left( \Gamma_k+\Gamma_l \right).
\label{full form M}
\en
for Type $M$. The successive inclusion of poles in the radicand of \eref{gams}, i.e. increasing numbers of $P_i \neq 0$, incrementally decreases the size of the corresponding commuting family. In particular, two such poles yields precisely the Type 2 matrices of Sect.~\ref{type2} (see in particular \eref{GamGam}), i.e. for $M=2$, the ansatz is precisely the complete parameterization of Type 2 given in the previous section.  

%\subsection{\label{Type I Rev}Type I/Maximal Operators -- Revisited}

For $M=1$, the ansatz offers an alternate, but equivalent parameterization of Type 1 matrices to that given in \esref{Hmel}. To see this, consider the off-diagonal elements of a general Type 1 matrix given by the ansatz in light of the equation $\Gamma_m^2-\Gamma_n^2=P_1(\varepsilon_m-\varepsilon_n)/( \lambda_1-\varepsilon_m ) ( \lambda_1-\varepsilon_n )$:
\beg
\left[H \left(u\right)\right]_{mn}=\gamma_m \gamma_n \dfrac{1}{2} \dfrac{d_{m}-d_{n}}{\Gamma_m-\Gamma_n} \dfrac{P_1}{( \lambda_1-\varepsilon_m ) ( \lambda_1-\varepsilon_n )}, \quad m \neq n\label{Type I-max}
\en
we can re-express \eref{Type I-max} in the \eref{Hmel} form, i.e.
\[
\left[H \left(u\right)\right]_{mn}= \widetilde{\gamma}_{m} \widetilde{\gamma}_{n}
\left(\frac{d_{m} -d_{n} }{\widetilde{\varepsilon}_{m}-\widetilde{\varepsilon}_{n}}\right),\quad m\ne n,
\]
if we have
\begin{equation}
\begin{array}{l}
\dis \widetilde{\varepsilon}_j=\frac{2}{P_1} \dfrac{w \Gamma_j+x}{y  \Gamma_j+z}\\
\\
\widetilde{\gamma}_j=\dfrac{\gamma_j}{\left( \lambda_1-\varepsilon_j \right) \left( y  \Gamma_j+z \right)},\\
\end{array}
\label{type I eqs}
\end{equation}
where $w,x,y,z$ are arbitrary real numbers satisfying $w z-x y=1$. Note that this constitutes a threefold redundancy in the ansatz parameterization of Type 1 matrices. In Sect.~\ref{Gauge} we will see that this redundancy is in addition to another threefold parametric redundancy that all ansatz matrices share. Demonstrating the equivalence of the diagonal elements of maximal matrices as defined by \esref{Hmel} to those of ansatz Type 1 matrices, as given by \esref{Hmel2}, proceeds similarly. 

The above is sufficient to show that  Type 1 matrices, as specified by the above ansatz, are precisely  the Type 1 matrices of Sec.~\ref{type1}. Note, however, that there is a subtlety in the definition of $d_{m}$. In particular, the ansatz seemingly stipulates (i.e. through \eref{dmM}) a restricted set of allowed $V$, even for Type 1,  whereas the \esref{Hmel} take the $V$  to be diagonal matrices with arbitrary real diagonal elements $d_m$. In fact, these are equivalent statements because for $M=1$ there are $N$ $g_j$ and $N$ $d_m$ in  \eref{dmM}  related via an $N\times N$ matrix. This matrix is similar to the Cauchy matrix\cite{cauchy} and can be similarly shown to be nondegenerate. Therefore, for an arbitrary choice of $d_m$ there always exists the corresponding choice of $g_j$ and vice versa.

Note that \esref{Hmel}, \re{Hmel2} and \re{gams}, $M=1$, are explicitly identical when $P_1=0$ and \emph{all} $\Gamma_m=1$. However, the ansatz allows for some $\Gamma$'s to be $-1$ when $P_1=0$. So for an ansatz Type 1 matrix with $P_1=0$ -- which we might call a `Type 0' matrix -- if $\Gamma_i=\Gamma_j$, $S_{ij}$ has the form \eref{S1}. If, however,  $\Gamma_i=1$ and $\Gamma_j=-1$, $S_{ij}$ vanishes. It follows that if we take a Type 1 commuting family and tune $|P_1|$ ever smaller, while keeping all other parameters fixed, we find that when $P_1=0$ we are left with pseudo-Type 1 matrices with block diagonal structure and, consequently, $u$-independent symmetry. Indeed, these `Type 0' commuting families will be the direct sum of two different Type 1 families.

\subsection*{\label{reality}Ensuring reality of ansatz matrices}

Next, we discuss the conditions on ansatz parameters such that the resulting matrices are real symmetric. For unrestricted complex parameters $\gamma_i$, $\varepsilon_i$, $P_i$, $B$ and $g_i$ the above construction, Eqs.~\re{Hmel2} and \re{gams},   produces complex symmetric matrices for which all properties of Type $M$ matrices save for reality, but including the exact spectra derived in the next section hold. 

Moreover, it can be shown that there exist complex parameters such that the corresponding matrices are real Type $M$, and where the complexity cannot be removed simply by multiplication by some common factor. A block of the 1d Hubbard model  in Appendix~\ref{block} provides an interesting example of such a situation. However, the most general constraint on {\it complex} parameters $\gamma_i$, $\varepsilon_i$, $P_i$, $B$ and $g_i$ to ensure real Type $M$ matrices appears to be rather non-trivial and we will not attempt to derive it here. Instead, let us choose the parameters $B,\gamma_j,\varepsilon_j,P_i$ and $g_i$ to be real and constrain them so that Eqs.~\re{Hmel2} and \re{gams} always yield real matrices. In this case the only problematic component for producing real matrices is the $\Gamma_j$ -- we still need  the radicand in  \eref{gams} to be positive.

First, we rewrite the expression under the square root in \eref{gams} as follows
\beg
1+\sum_{j=1}^{M} \frac{P_j}{\lambda_j-\sigma} = \dfrac{\prod_{j=1}^{M}{(\phi_j-\sigma)}}{\prod_{j=1}^{M}{(\lambda_j-\sigma)}},   
\en
i.e. we traded $M$ parameters $P_j$ for new $M$ parameters $\phi_j$. Note that $P_j$'s can be expressed in terms of $\phi_j$'s by computing the residues at the corresponding poles
$$
P_j=\frac{\prod_{k=1}^M(\phi_k-\lambda_j)}{\prod_{k\ne j}(\lambda_k-\lambda_j)}
$$

Let us order $\varepsilon_j$ so that $\eps_j < \eps_{j+1}$. 
Recall that $\lambda_j$ are solutions of \eref{Gaudin1}. It can be shown (see e.g. Ref.~\onlinecite{owusu}) that they are all real and lie between consecutive $\eps_j$. We have $\eps_i<\lam_i<\eps_{i+1}$ for $B>0$ and $\eps_{i-1}<\lam_i<\eps_i$ for $B<0$, where $i=1,\dots,N$.   If we now take $\phi_j$, $j=1,\dots,M$, to be also real  and such that
\beg
\begin{array}{l}
\eps_j< \phi_j<\lam_j\quad \mbox{for $B>0$}\\
\lambda_j < \phi_j < \varepsilon_j\quad \mbox{for $B<0$}\\
\end{array}
\en
all terms $\frac{\phi_j-\varepsilon_j}{\lambda_j-\varepsilon_j}>0$, which is sufficient (though not necessary) to ensure the positivity of the radicand in Eq.~\ref{gams} and reality of the matrix elements.

 \section{\label{Exact}Exact solution}
 
In this section, we present the exact solution of the Type $M$ matrices parameterized by the above ansatz, see Sec.~\ref{Type M}. The exact solution of the eigenvalue problem for Type $M$ matrices proceeds in a manner reminiscent of that of the Type 1 operators of Ref.~\onlinecite{owusu}, see also Sec.~\ref{type1}. A particularly striking feature of this exact solution is that, though these operators have explicitly more structure than Type 1 operators, the exact solution is nevertheless essentially just as simple, in that the entire eigensolution derives directly from the solution of a \emph{single} algebraic equation. This is in stark contrast to the situation typical with Bethe's ansatz where there are a large number of coupled algebraic equations to be solved to derive a small portion of the spectrum.

 We want to determine the exact solution $H^i (u) \cdot \vec{v}^\sigma (u)=E_\sigma^i(u) \vec{v}^\sigma (u)$, where $H^i(u)$ is a basic operator specified in \eref{him},  $\vec{v}^\sigma (u)$ and  $E_\sigma^i(u)$ are its eigenvectors and corresponding eigenvalues, respectively.  The ansatz eigensolution is as valid for Type 1, as it is for all ansatz Type $M$ and so the components of eigenvector $\vec{v}^\sigma (u)$ must bear some resemblance to that of our existing Type 1 exact solution. Let us modify \eref{eigenvector1} somewhat and assume that the $j^{\mbox{{\scriptsize th}}}$ component of this eigenvector is given by 
  \beg
v^\sigma_j \equiv \left[\vec{v}^\sigma (u)\right]_j= \dfrac{1}{2} \dfrac{\gamma_j}{\sigma-\varepsilon_j} \left( \Gamma(\sigma) + \Gamma_j\right).
\label{eigenvector}
\en
Expanding in terms of our parameters and using Eqs.~\re{him} and \re{full form M},  we obtain, after a bit of algebraic manipulation
\[
\left[H^i \cdot \vec{v}^\sigma\right]_{j}=v^\sigma_j \sum_m{\dfrac{1}{\lambda_i-\varepsilon_m} \dfrac{\gamma_m^2}{2} \dfrac{\Gamma(\sigma)+\Gamma_m}{\sigma-\varepsilon_m}}+\dfrac{v^\sigma_j}{\lambda_i-\varepsilon_j} \left[ u- \sum_m{\dfrac{\gamma_m^2}{2} \dfrac{\Gamma(\sigma)+\Gamma_m}{\sigma-\varepsilon_m}}-\dfrac{B}{2} \bigl( \Gamma(\sigma)-1\bigr)   \right]
\]
Note, the second term is a $j$-dependent coefficient on the $j^{th}$ putative eigenvector component. This term must vanish in order for \eref{eigenvector} to constitute an eigenvector. 
Thus, given an eigenvector of matrix $H^i(u)$ specified by \eref{eigenvector} the corresponding eigenvalue is
\beg
E_\sigma^i=\sum_m{\dfrac{1}{\lambda_i-\varepsilon_m} \dfrac{\gamma_m^2}{\sigma-\varepsilon_m} \dfrac{1}{2} \big(\Gamma_m+\Gamma(\sigma) \big)},
\label{eigened}
\en
where $\sigma$ is such that the following algebraic equation
\beg
 \sum_{m}{\dfrac{1}{2} \dfrac{\gamma_m^2}{\sigma-\varepsilon_m} \left(\Gamma(\sigma) + \Gamma_m \right)}-\dfrac{B}{2}\left( \Gamma(\sigma)-1\right)=u,
 \label{constraint}
\en
 is satisfied. We should note, however, that it remains to be proven that \eref{constraint} generically yields $N$ \emph{distinct} solutions given both sign choices for $\Gamma(\sigma)$, see \eref{gams}. Nevertheless, numerical analysis has yielded no counterexample.

\begin{figure}[hpt]
\begin{center}
\includegraphics[scale=1.2]{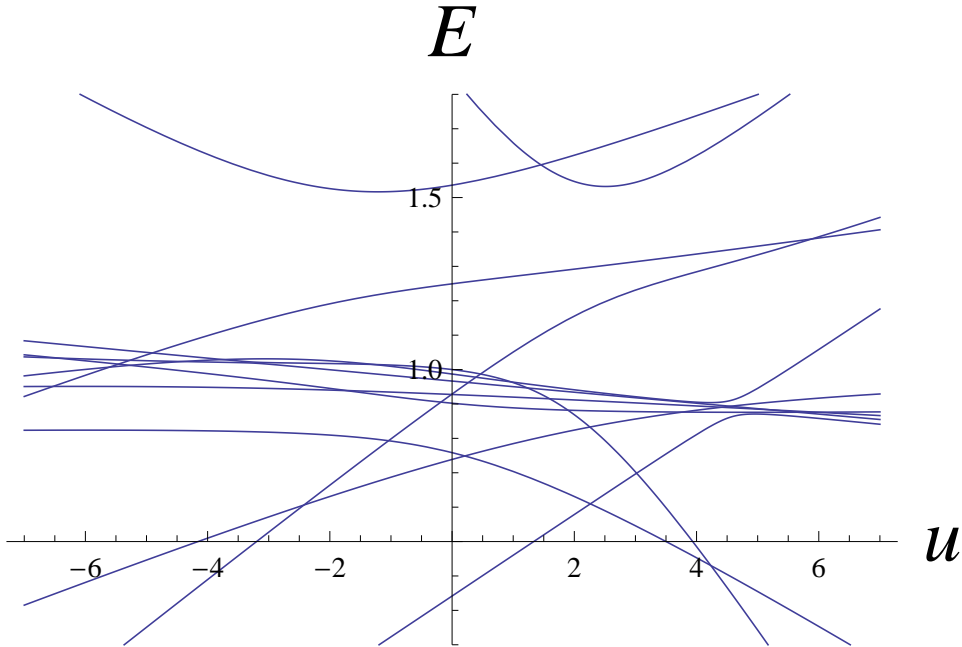}  
\caption{\label{plotty} Numerical energy levels of a  $12\times12$ Type 4 operator $H(u)=\sum_{i=0}^{8}{g_i H^i(u)}$ with parameters $B,\gamma_j,\varepsilon_j, P_j$, and $g_j$ randomly chosen over a uniform distribution of real numbers. The eigenvalues exhibit many level crossings, frequently violating the Wigner-von Neumann non-crossing theorem.}
\end{center}
\end{figure}

\begin{figure}[hpt]
\begin{center}
\includegraphics[scale=0.8]{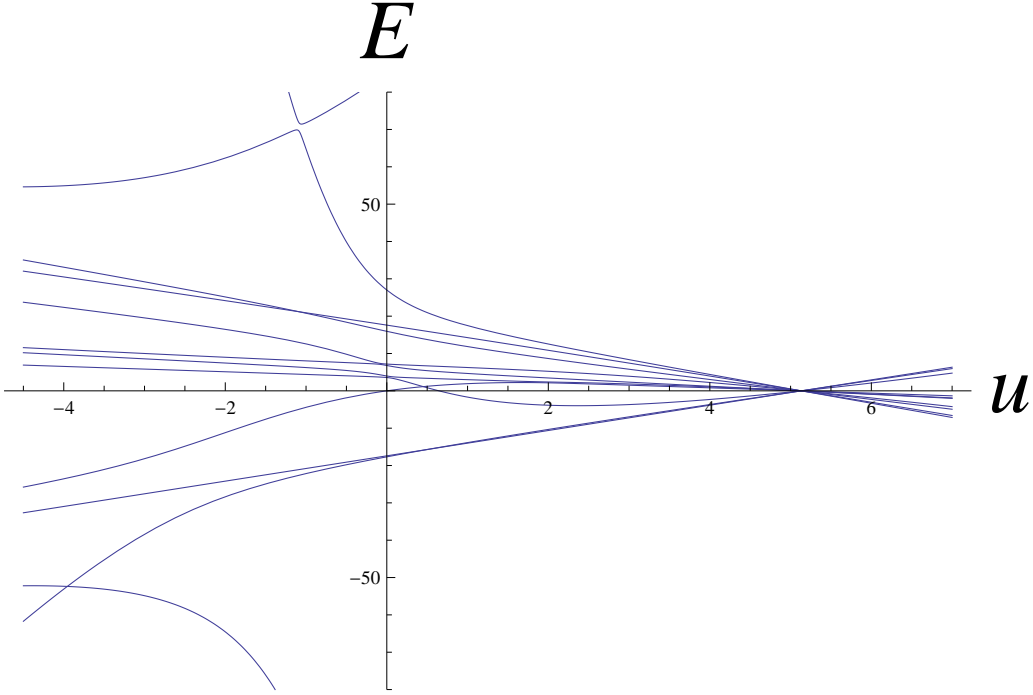}  
\caption{\label{plotty2} Numerical energy levels of a  $12\times12$ Type 4 operator $H^i(u)$, $i=3$, with parameters $B,\gamma_j,\varepsilon_j,\lambda_i,P_i$ identical to those in Fig.~\ref{plotty}. Note that the eigenvalues exhibit $N-2=10$ concurrent level crossings at a particular value of $u$.}
\end{center}
\end{figure}

The behavior of the spectrum of a typical Type $M$ operator as a function of the parameter $u$ is particularly interesting. As distinct from  the spectra of random real symmetric operators linear in a parameter, random Type $M$ matrices -- whose parameters $B,\gamma_j,\varepsilon_j,\lambda_i,P_i$ are chosen from random uniform distributions -- exhibit frequent violations of the Wigner-von-Neumann non-crossing rule, see Fig.~\ref{plotty}, which states that eigenvalues of the same symmetry do not cross as a function of a single coupling parameter. 

It is interesting to note that numerical observation of individual $H^i(u)$ suggests that they undergo concurrent crossings of $N-2$ eigenvalues at some value of $u$, see Fig.~\ref{plotty2}. This is a surprising phenomenon that one would rightly expect from Type 2 matrices \cite{type2} but, perhaps, not general Type $M$ matrices. 

We can explain this phenomenon by considering the $H^i(u)$ eigenvalues. At a given $u$, and for all $\sigma$ satisfying \eref{constraint} we find, through some algebraic manipulation
\beg
E_\sigma^i=\sum_m{\dfrac{1}{\lambda_i-\varepsilon_m} \dfrac{\gamma_m^2}{\sigma-\varepsilon_m} \dfrac{1}{2} \big(\Gamma_m+\Gamma(\sigma) \big)}=\dfrac{1}{\lambda_i-\sigma} \left[u-\dfrac{1}{2}\biggl(B+\sum_{m}{\dfrac{\gamma_m^2}{\lambda_i-\varepsilon_m}\Gamma_m} \biggr) \right].
\en
Thus, at $\widetilde{u}=\frac{1}{2}\bigl(B+\sum_{m}{\frac{\gamma_m^2}{\lambda_i-\varepsilon_m}\Gamma_m} \bigr)$ we find that, for $\sigma \neq \lambda_i$, the $\sigma$'th eigenvalue of $H^i(\widetilde{u})$ is zero. Naturally, not all the eigenvalues can vanish, otherwise $H^i(\widetilde{u})$ is trivial. 
Indeed, setting $\sigma=\lam_i$ and $u=\widetilde{u}$ in \eref{constraint}, we find that it is solved for both sign choices for $\Gamma(\sigma)$, i.e.  \eref{constraint} has a doubly degenerate solution $\sigma_-(\widetilde{u}) =\sigma_+(\widetilde{u})= \lambda_i$, such that $\Gamma(\sigma_+)=-\Gamma(\sigma_-)$. 
Numerically, we find confirmation of this phenomenon wherein $N-2$ $\sigma$'s correspond to zero eigenvalues, while the other two eigenvalues at $\widetilde{u}$ are non-zero.

 \section{\label{Gauge}Gauge redundancy in the ansatz parameters}

Given a set of real parameters $B,\gamma_j,\varepsilon_j,\lambda_i,P_i$ defining a Type $M$ commuting family one might ask whether these parameters are unique, i.e. does there exist a set $\widetilde{B},\widetilde{\gamma}_j,\widetilde{\varepsilon}_j,\widetilde{\lambda}_i,\widetilde{P}_i$ yielding the same Type $M$ commuting family?

On inspection it is apparent that parameters are, in fact, not unique,
e.g. a
 uniform shift of the $\varepsilon$ (and $\lambda$) parameters leaves Eqs.~\re{Hmel2} and \re{gams} unchanged. 
 Similarly, one can find a variety of uniform scaling schemes under which all matrix elements are invariant. 

More interesting, however, is the following linear fractional parameter transformation:
\beg
\widetilde{\varepsilon}_j=\dfrac{w\varepsilon_j+x}{y \varepsilon_j+z}, \qquad
\widetilde{\lambda}_j=\dfrac{w\lambda_j+x}{y\lambda_j+z}, \qquad
\widetilde{\gamma}_j=\dfrac{\gamma_j}{y \varepsilon_j+z} \Gamma(\varsigma),  \qquad
\widetilde{P}_j=\frac{1}{\Gamma^2(\varsigma)} \dfrac{P_j}{(y \varepsilon_j+z)^2},
\label{transformer}
\en
where
$w,x,y,z$ are arbitrary real parameters subject to constraint $w z-x y=1$, $\varsigma= -z/y$ and $\Gamma(\varsigma)$ is defined in \eref{gams}. It can be shown then that the above transformation yields the corollary transformation:
\beg
\widetilde{\Gamma}_j=\dfrac{\Gamma_j}{\Gamma(\varsigma)}.
\en

By direct computation it can be shown that the $S_{ij}$ and, therefore, all off-diagonal matrix  elements of the members of the ansatz commuting families are invariant under the above. Now consider the action of this transformation on diagonal matrix elements of basic commuting matrix $H^i(u)=T^i+uV^i$ given by Eqs.~\re{him} and \re{vibasis}. For $V^i$ we have
\beg
\widetilde{d}_j^i \equiv \Big( \dfrac{\lambda_i}{1-\varsigma^{-1} \lambda_i}-\dfrac{\varepsilon_j}{1-\varsigma^{-1}\varepsilon_j} \Big)^{-1}=d_j^i (1-\varsigma^{-1} \lambda_i)^2+\varsigma^{-1} (1-\varsigma^{-1} \lambda_i).
\en
We see that the effect of the transformation on diagonal matrix $V^i$ is a uniform scaling by $(1-\varsigma^{-1} \lambda_i)^2$ and shift by a trace, $\varsigma^{-1}(1-\varsigma^{-1} \lambda_i) I$, i.e. 
\beg
\widetilde{V}^i=(1-\varsigma^{-1} \lambda_i)^2 V^i+\varsigma^{-1}(1-\varsigma^{-1} \lambda_i) I.
\label{V transform}
\en

Using partial fraction decomposition, see \eref{Gaudin sep}, and \eref{gams} it can be shown that 
$$
(1-\varsigma^{-1} \lambda_i)^{-2}\widetilde{H}^i(u)= H^i\biggl[\dfrac{u}{(1-\varsigma^{-1} \lambda_i)^2}+ \sum_{j}{ \gamma_{j}^{2} \dfrac{\Gamma(\varsigma)+\Gamma_j}{\varsigma-\varepsilon_j}}-\dfrac{B}{2} \left( \Gamma(\varsigma)-1 \right)  \biggr] -I \sum_{j}{ \gamma_j^2\dfrac{\Gamma_j+\Gamma(\varsigma)}{(\lambda_i-\varepsilon_j)(\varsigma-\varepsilon_j)}},
$$
i.e. the gauge transformation preserves the matrix elements of the Type $M$ matrices, up to a uniform rescaling, a shift in the coupling parameter $u$ and the addition of a multiple of the identity matrix.

\subsection*{\label{2gauge}Type 2}

With respect to the ansatz parameterization, in addition to the invariance under the above transformation, Type 2 matrices have a parametric redundancy not shared by Types $M>2$. Numerical work undertaken to ``reverse engineer'' ansatz parameters directly from the matrix elements of $M>2$ commuting matrices  produced by the ansatz (see Appendix~\ref{inverse}) yield back the input parameters, modulo the transformation \re{transformer}. However, Type 2 matrices exhibit a 1-dim gauge freedom beyond that of \esref{transformer}.

To see where this extra redundancy comes from, consider again \eref{tilde form}. Let us now allow $\varepsilon_m$ to be a smooth function of a parameter $t$ such that
\beg
\widetilde{S}_{mn}=\dfrac{1}{2} \dfrac{\widetilde{\Gamma}\left( \varepsilon_m (t),\varepsilon_n (t) \right)+\widetilde{\Gamma} \left( \varepsilon_n (t),\varepsilon_m (t) \right)}{\varepsilon_m(t)-\varepsilon_n(t)},
\label{tilde form2}
\en
where $\widetilde{\Gamma} \left( \alpha,\beta \right) \equiv \delta_m \sqrt{\left( \lambda_1-\alpha \right) \left( \lambda_2-\alpha \right) \left( \phi_1-\beta \right) \left( \phi_2-\beta \right)}$ and $\phi_1,\phi_2$ are defined by the equation 
\[
\frac{\left( \phi_1-\alpha \right) \left( \phi_2-\alpha \right)}{\left( \lambda_1-\alpha \right) \left( \lambda_2-\alpha \right)}=1+\frac{P_1}{\lambda_1-\alpha}+\frac{P_2}{\lambda_2-\alpha}
\]
for all $\alpha$. By direct computation it can be shown that $\widetilde{S}_{mn}$ is $t$-independent if 
\beg
\left( \dfrac{d\varepsilon_m}{dt} \right)^2=\left( \lambda_1-\varepsilon_m(t) \right) \left( \lambda_2-\varepsilon_m(t) \right) \left( \phi_1-\varepsilon_m(t) \right) \left( \phi_2-\varepsilon_m(t) \right),
\label{inv}
\en 
i.e.  $\varepsilon_m(t)$ is the elliptic function of $t$ corresponding to \eref{inv} with initial conditions $\varepsilon_m (0) \equiv \varepsilon_m$. Thus  $\widetilde{S}_{mn}$ is invariant under the transformation $\varepsilon_m (0) \rightarrow \varepsilon_m (t)$.

A general elliptic function -- i.e. one defined by \eref{inv} wherein $\lambda_1,\lambda_2,\phi_1,\phi_2$ are arbitrary complex numbers -- is related to a specific Jacobi elliptic function through a linear fractional transformation. Concretely, 
\beg
\varepsilon_j(t)=\dfrac{w k \, \mathrm{sn}\left( (t_j+t)/\tau \right)+x}{y k\, \mathrm{sn} \left( (t_j+t)/\tau \right)+z},
\label{linfrac}
\en
where   $w,x,y,z$ are complex numbers such that $w z-x y=1$, and
\beg
\dfrac{w \lambda_1+x}{y \lambda_1+z}=k^{-1}, \quad \dfrac{w \lambda_2+x}{y \lambda_2+z}=-k^{-1},\quad \dfrac{w \phi_1+x}{y \phi_1+z}=k \quad \dfrac{w \phi_2+x}{y \phi_2+z}=-k,
\en
$\tau=\pm k \sqrt{\left( -y^2 k^2+w^2 \right) \left( -y^2 k^{-2}+w^2 \right)}$, and $k$ is any one of the four roots of the equation 
\beg
\left( \frac{k-k^{-1}}{k+k^{-1}} \right)^2=\frac{(\lambda_1-\phi_1)(\lambda_2-\phi_2)}{(\lambda_1-\phi_2)(\lambda_2-\phi_1)}.
\en

Elliptic functions have a number of beautiful transformation properties. Interestingly, it can be shown, using the well known `angle' addition formulae involving Jacobi elliptic functions\cite{Abramowitz} that
\beg
\mathrm{sn}(t/\tau+\kappa/4)=\Gamma_j/\Gamma \left(-z/y \right),
\en
where $\Gamma(\sigma)$ is defined as in \eref{GamGam} and $\kappa$ is one of the characteristic periods of these doubly-periodic functions. This is a surprising intermingling of $\varepsilon_j$ and $\Gamma_j$, the broader meaning of which remains mysterious. 

The most compelling such exploitation of the behavior of Jacobi elliptic functions comes in reconsidering \eref{tilde form2}. By directly substituting \eref{linfrac} into \eref{tilde form2} we find that
\beg
\widetilde{S}_{mn}=\dfrac{1}{2 \tau} \dfrac{\mathrm{cn}\left((t_m+t)/\tau \right) \mathrm{dn}\left((t_n+t)/\tau \right)+\mathrm{cn}\left((t_n+t)/\tau \right) \mathrm{dn}\left((t_m+t)/\tau \right)}{\mathrm{sn}\left((t_m+t)/\tau \right)-\mathrm{sn}\left((t_n+t)/\tau \right)}.
\en
However $\widetilde{S}_{mn}$ is independent of elliptic parameter $t$. We can make this $t$-invariance manifest by using the addition law properties of Jacobi elliptic functions\cite{Abramowitz}, whereby it  can be shown that 
\beg
\widetilde{S}_{mn}=\dfrac{1}{2 \tau} \dfrac{\mathrm{cn}\left(\dfrac{t_m-t_n}{2\tau} \right)\mathrm{dn}\left(\dfrac{t_m-t_n}{2\tau} \right)}{\mathrm{sn}\left(\dfrac{t_m-t_n}{2\tau}\right)},
\label{ellS}
\en
where $\mathrm{sn}^2+\mathrm{cn}^2=1$ and $k^4 \mathrm{sn}^2+\mathrm{dn}^2=1$.

This parametric redundancy is peculiar to Type 2 -- the naive generalization to higher types, i.e. extending Eqs.~(\ref{tilde form2}) and \re{inv} by including $\lambda_i,\phi_i$, $1 \leq i \leq M$ and generating $\varepsilon_j(t)$'s that satisfy hyperelliptic differential equations does not yield an invariant analogue of \eref{tilde form2}. That such a simple, geometric redundancy is embedded in the derived construction of Type 2 matrices is both beautiful and compelling, however its \emph{meaning} is unclear.

\section{\label{numerics}Random generation of commuting matrices and completeness of the ansatz }

In this section we address the issue of completeness of the ansatz --  whether \emph{all} Type $M>2$, commuting matrices are \emph{also} ansatz matrices, i.e. for any pair of commuting matrices both linear in a parameter $u$, do ansatz parameters $B,\gamma_j,\varepsilon_j,\lambda_i,P_i$ and $g_i$ exist that yield back these matrices? As yet, we have no analytic means to directly answer this question, but numerically we find that the answer is `\emph{Yes}' for $M=3$  and `\emph{No}' for $M>3$. Specifically,  numerical analysis shows that $2N+2M-5$ continuous parameters are needed to uniquely specify a generic Type $M$ commuting family, i.e. the ansatz of Sect.~\ref{Type M} is short by $M-3$ parameters. 

Below we will detail this analysis, wherein we

\begin{enumerate}
\item generate two random commuting matrices and determine the type of the commuting family they belong to,

\item then process one of these matrices through an algorithm designed to check whether  a matrix belongs to an ansatz Type $M$ commuting family and extract ansatz parameters if it does.

\end{enumerate}

Recall from Sect.~\ref{strategy} that  any two commuting symmetric matrices $H^1(u)=T^1+uV^1$  and $H^2(u)=T^2+uV^2$  can be represented as $H^1(u)=u V^1+[V^1,S]+W^1$ and $H^2(u)=u V^2+[V^2,S]+W^2$, where $V^1,V^2,W^1,W^2$ are diagonal and $S$ is an antisymmetric matrix. $[H^1(u), H^2(u)]=0$ is then equivalent to (see \eref{comm2})
\beg
\left[ [V^1,S], [V^2,S] \right]+ \left[[V^1,S],W^2 \right]-\left[[V^2,S],W^1 \right]=0.
\label{comm3}
\en
It turns out that we can generate a pool of generic commuting matrices by brute forcing a solution to \eref{comm3}. Using  Mathematica's {\it FindRoot}\cite{wolf} function -- an algorithm designed to search a $D$-dimensional parameter space for solutions to a set of $D$ equations -- we input $4N$ random real values for the nonzero elements of $V^1,V^2,W^1,W^2$ and ask  FindRoot to find a solution $S$ to \eref{comm3} whose matrix elements are close to those of a randomly generated antisymmetric seed matrix. Typically  FindRoot is able to quickly find such an antisymmetric matrix. This solution appears to be one of many in a large discrete set of compatible antisymmetric matrices, not all of them real. We know that there are many solutions because changing the random seed matrix, given the same inputs $V^1,V^2,W^1,W^2$, frequently sees  FindRoot land on an entirely different antisymmetric matrix. 

We believe the solution set discrete because when Mathematica's \emph{NSolve} -- an algorithm designed to search for \emph{many} numerical solutions to an arbitrary set of constraints -- attempts to solve  \eref{comm3} given those same $4N$ inputs it manages to generate a discrete set of $S$. However if we decrease the number of inputs, e.g. specify only $4N-1$ inputs,  NSolve very quickly indicates that the system is under-constrained by at least one equation and randomly generates a linear constraint so as to proceed toward calculating a solution set.

In practice, we observe that  FindRoot algorithm can find at least one solution to \eref{comm3}, given $4N$ inputs. Notice, though, that this method of randomly generating commuting matrices is  independent of matrix type, i.e. nowhere in the procedure does one explicitly specify the size of the commuting family to which $H^1(u)$ and $H^2(u)$ belong. The initial $4N$ parameters defining matrices $V^1,V^2,W^1,W^2$ and the antisymmetric solution matrix $S$ correspond to some Type $M$ commuting family, and we can determine the value of $M$ by counting the number of linearly independent matrices in the commuting family. 

In particular we take a putative matrix $H^k(u)=u V^k +W^k +[V^k,S]$, where $V^k$ and $W^k$ are diagonal matrices whose nonzero entries, $d^k_m$ and $f^k_m$ respectively, are undetermined real variables, and solve commutation relation $[H^1(u),H^k(u)]=0$. There will be $N(N-1)/2$ equations linear in these $2N$ unknown $d^k_m$ and $f^k_m$. Let $p$ be the number of these equations found to be linearly independent. Then it follows that, because $H^k(u)$ is a Type $M$ matrix and, by definition, we can only freely choose $K=N-M+1$ of the non-zero matrix elements of $V^k$ and the
overall trace of $W^k$, $p=2N-(K+1)$, i.e. the commutator yields $p=N+M-2$ linearly independent constraints on $V^k$ and $W^k$. Thus counting the number $p$ of independent equations tells us the matrix type, $M=p-N+2$.

\begin{figure}[h]
\begin{centering}
\includegraphics[width=.96\textwidth]{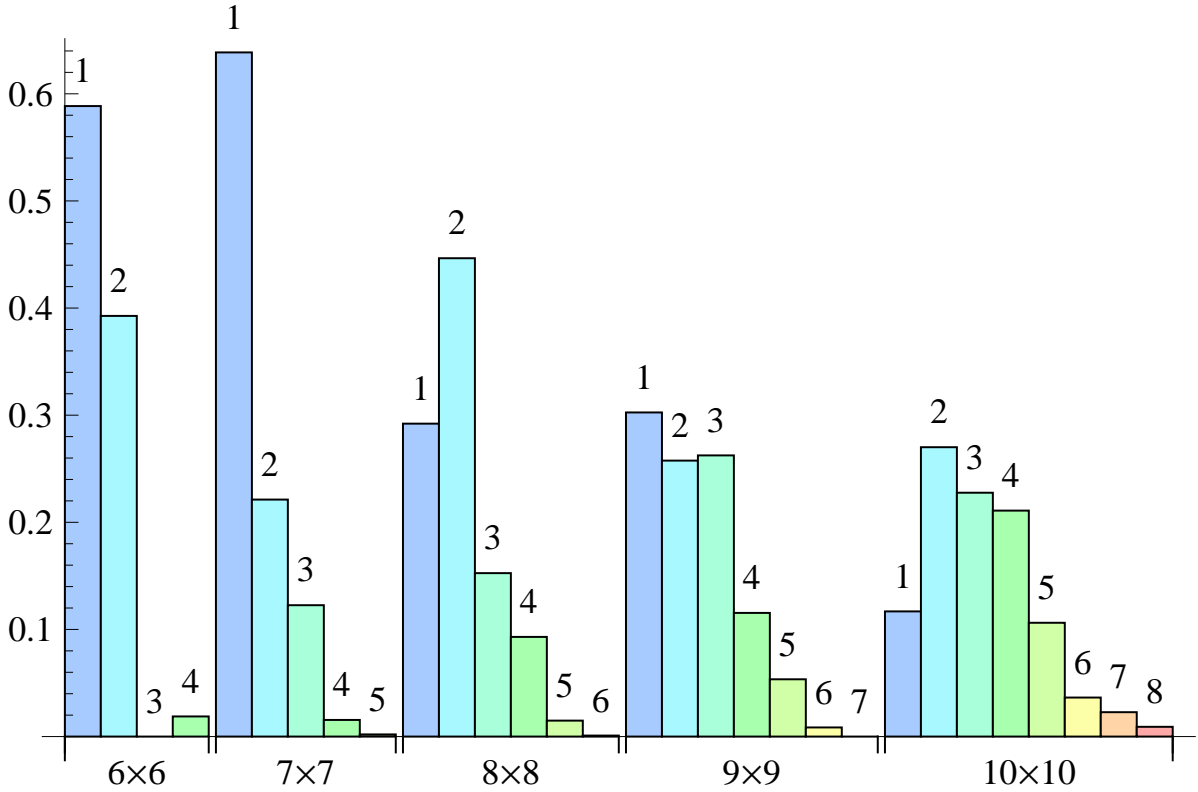}
\end{centering}

\caption{\label{table}  The distribution of various matrix types $1 \leq M \leq N-2$ among randomly generated integrable matrices of sizes $N=6,7,8,9$ and 10.  $4N$ random real values (flatly distributed between $-1$ and $1$) are given to the nonzero matrix elements of diagonal matrices  $V^1,V^2,W^1,W^2$ and \eref{comm3} is solved numerically. At each iteration of random generation, one such $S$ solution is chosen among a large set of such, and matrices $H^1(u)$ and $H^2(u)$ are constructed and their type determined. We observe that random generation favors lower types, and we note with some surprise that there are types that appear with extremely low (albeit finite) frequency, e.g. $8 \times 8$ Type 6, or not at all, $6 \times 6$ Type $3$.}
\end{figure}

We observe that this procedure for generating generic commuting matrices yields a distribution of types weighted toward the lower end, see Fig.~\ref{table}. Interestingly,  out of many hundreds of random generations, not one $6 \times 6$ Type 3 matrices was generated, even though we know that they exist and can be readily made by the ansatz.

We cannot as yet prove that given some random diagonal matrices $V^1,V^2,W^1,W^2$ a solution $S$ must always exist for \eref{comm3}, but for the reasons stated above we find empirically that $4N$ parameters are always sufficient and apparently necessary to generate two distinct commuting matrices. Accordingly, let us assume $4N$ the correct number of parameters necessary to specify two distinct, generic $N \times N$ real symmetric commuting matrices, linear in parameter $u$, regardless of type. If those two matrices are Type $M$, of those $4N$ inputs $2(N-M+2)$  are used to uniquely determine the two independent members of the commuting family, i.e. $2(N-M+1)$ for both unique $V$'s and an extra $2$ inputs to fix a trace on both $W$'s not constrained by the commutation relations. If, in addition, we remove an overall scale on $S$ matrix elements, that leaves $2N+2M-5$ of our initial $4N$ input parameters that remain to uniquely specify a generic Type $M$ commuting family.

\subsection*{\label{status} Are all Type $M$ matrices also ansatz?}

We already know that all Type $1$ and $2$ matrices conform to the ansatz -- this is essentially what the ansatz was designed to do, see Sects.~\ref{type2} and \ref{Type M}. Now let us consider ansatz Type $M\ge 3$ matrices. Counting ansatz parameters $B,\gamma_j,\varepsilon_j,P_i$ and removing the parametric redundancies (see Sect.~\ref{Gauge}) we are left with $2N+M-2$ parameters to uniquely specify an ansatz Type $M \geq 3$ family. Comparing to the number of parameters needed to specify a generic Type $M$ commuting family, we see that ansatz Type $M$ families appear smaller  by $M-3$ continuous parameters.

These parameter counting arguments strongly suggest that a generic Type $M$ commuting matrix is not an ansatz matrix, i.e. the ansatz generates only a small subset of Type $M$ for $M>3$. The case of Type 3 appears marginal, i.e. on the one hand the preceding argument indicates that there are sufficient parameters for ansatz to cover all of Type 3, on the other hand the ansatz is a naive extension on a full parameterization of Type 2 and there is no obvious reason for the ansatz to extend fully to Type 3.

To gain further insight into the completeness of the ansatz, we  randomly generate generic commuting matrices of arbitrary type (see above) and subject them to an algorithm (described in detail in Appendix~\ref{inverse}) that reverse engineers the ansatz parameters from the matrix elements if the matrix is ansatz Type $M$. 
However, when we run generic commuting matrices through this algorithm, all observed matrices of Type $M>3$ are seen not to be ansatz matrices. In particular, in these cases the algorithm is unable to find nontrivial $\Gamma_j$ satisfying \eref{xi}. 

What is surprising, though, is that the ansatz carries no guarantee that it covers Type $3$, and yet \emph{every} observed, randomly generated Type $3$ matrix has an ansatz parameterization. In view of this result, we conjecture that the ansatz is complete for Types $1$ and $2$ -- which we have proved -- \emph{and} for Type $3$, while incomplete for general Type $M>3$ by an unknown set of $M-3$ parameters, which we note empties for Type $3$. In Sect.~\ref{discussion} we provider further arguments supporting this conjecture.

\section{\label{xings}Level crossings}

\begin{figure}[hpt]
\begin{center}
\includegraphics[scale=0.5]{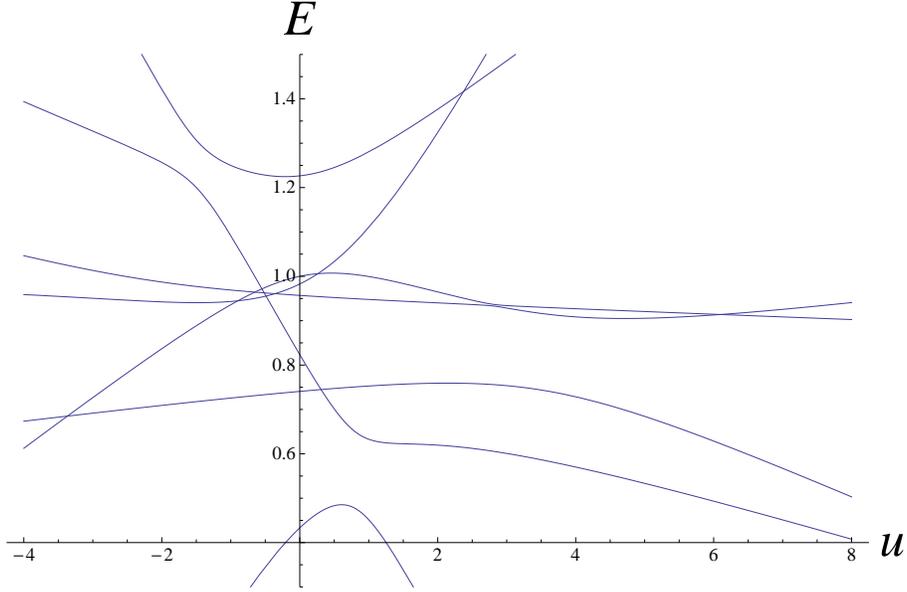} %\includegraphics[scale=0.8]{plot.eps}
\caption{\label{violator} The $u$-dependent spectrum of a randomly generated $7\times 7$ Type 4 operator $H(u)$. Generally the discriminant of the characteristic polynomial of an $N \times N$ matrix $T+uV$ is a polynomial in $u$ of order $N(N-1)$. We observe that discriminants of Type $M$ matrices factor as $\mathrm{Disc}_{H(u)}=P(u)Q(u)^2$. While the complex roots of $P(u)$ correspond to values of $u$ for which the spectral decomposition of $H(u)$ involves Jordan blocks, the crossings occur at roots of the $(N-1)(N-2)-M+1=12$ order polynomial $Q(u)$. This Type 4 matrix has 10 crossings at $u=\{-3.37, -0.90, -0.65, -0.55, -0.52, -0.49, -0.36, 0.31, 2.37, 6.10\}$ and we numerically confirm that they are roots of $Q(u)$. Its other 2 roots are complex conjugate, corresponding to `complex crossings' occurring at $u=\{0.61+1.95 i, 0.61-1.95 i\}$}
\end{center}
\end{figure}

One of the signatures of parameter dependent integrable models is the presence of level crossings in their spectra. A general $u$-dependent Hamiltonian, $H(u)$, obeys the Wigner-von Neumann non-crossing rule which states that when one plots the $u$-dependence of those energy levels sharing the same quantum numbers, the levels may approach one another closely, but they never cross\cite{Hund,Neumann,Teller,Landau,Longuet-Higgins,Naqvi,Kestner}. The spectra of Type $M$ Hamiltonians, however, behave very differently -- they carry frequent (though not necessary\cite{owusu}) violations of the non-crossing rule, see Figs.~\ref{plotty}, \ref{plotty2}, and \ref{violator}. 

For Type 1, we were able to offer a topological rationale for the necessary existence of at least one such crossing\cite{owusu}, but the analysis in Type $M$ is more difficult. We do observe, however that the number of crossings varies with type as 
\beg
\# \; \mbox{of crossings}=\dfrac{(N-1)(N-2)}{2}-\widetilde{M}+1-2L,
\label{Qq}
\en 
where $L$ is some non-negative integer bounded such that the above expression is non-negative, and where $\widetilde{M}$ is an integer such that
\[
\widetilde{M} \geq M.
\]
In the overwhelming number of observed commuting matrices $\widetilde{M}=M$. Specifically, out of several hundreds of matrices analyzed, only two non-ansatz matrices were observed to be such that $\widetilde{M} \neq M$. $\widetilde{M}$ then is overwhelmingly identical to type, but there are apparent exceptions (see Sect.~\ref{discussion}). Note, we always find $\widetilde{M} = M$ for ansatz matrices.

The crossings'  story gets more interesting when one extends the notion of crossings to ``complex crossings'', by which we mean complex values of $u$ wherein at least two of the eigenvalues of an operator $H(u)$ are equal. More concretely, complex crossings occur at values of $u$ corresponding to the roots of the discriminant of the characteristic polynomial ${\cal P}(\lam, u)$ of $H(u)$. The discriminant is defined as $\prod_{i<j}{(E_i-E_j)^2}$, where $E_i$ are the roots of ${\cal P}(\lam, u)$, i.e. it is polynomial in the coefficients of ${\cal P}(\lam, u)$ and, therefore, polynomial in $u$.  The order of this polynomial can be shown to be  $N(N-1)$, and, therefore, there are exactly $N(N-1)$ complex values of $u$ (roots) where the discriminant vanishes, independent of type or even whether $H(u)$ belongs to a commuting family. 

However, from numerical analysis we see that for a general $H(u)$ these roots do not correspond to  eigenvalues being equal one another; they correspond to a situation in which there ceases to be a full eigen-decomposition of the operator. For a completely general, non-integrable matrix, the roots of the discriminant correspond to a spectral decomposition of $H(u)$ involving Jordan blocks, rather than eigenspaces, see Ref.~\onlinecite{Horn}. A Type $M$ matrix, $H^i(u)$, however, is rather special. Its discriminant -- which we denote $\mathrm{Disc}_{H^i(u)}$ -- factorizes as
\[
\mathrm{Disc}_{H^i(u)}=P(u)Q^i(u)^2,
\]
where $P(u)$ and $Q^i(u)$ are polynomials. It turns out that $Q^i(u)$ is a polynomial of order $(N-1)(N-2)/2-\widetilde{M}+1$ and its roots are often real, but not always. Its complex roots come in complex conjugate pairs, hence the $2L$ term in \eref{Qq}. All of its roots correspond to values of $u$ where at least two eigenvalues are equal. Moreover, if we consider two Type $M$ matrices, $H^i(u)$ and $H^j(u)$ belonging to the same commuting family, i.e. $[H^i(u),H^j(u)]=0$, their discriminants will have distinct factors $Q^i(u)$ and $Q^j(u)$. However, they will have the same factor $P(u)$ and its $2N+2\widetilde{M}-4$ complex roots correspond to values of $u$ where a breakdown of the eigen-structure to Jordan blocks occurs. 

Interestingly we also observe that if we express $H^j(u)$ as a polynomial in $H^i(u)$ -- which we can do for any pair of non-degenerate commuting matrices, see Ref.~\onlinecite{Baum} -- the coefficients of that polynomial are rational functions in $u$ the denominators of which are \emph{all} $Q^i(u)$, i.e.
\beg
H^j(u)=\sum_{m=0}^{N-1}{\dfrac{R^j_m(u)}{Q^i(u)} \left[ H^i(u) \right]^m},
\en
where $R^j_m(u)$ are polynomials in $u$ of order $(N-1)(N-2)/2-\widetilde{M}-m+2$. Note that the failure of polynomials of a degenerate matrix to span that same matrix' commutant is reflected in the fact that the above polynomial's coefficients blow up at the roots of $Q^i(u)$, i.e. at those values of $u$ where $H^i(u)$ undergoes a level crossing.

\section{\label{Taxonomy}Taxonomy of types in the 1d Hubbard model}

In Sect.~\ref{numerics} we explored the distribution of general and ansatz matrix types among randomly generated commuting matrices. Here we study the prevalence of types in a well-known integrable system. As mentioned in the introduction (Sect.~\ref{Intro}), sectors of such systems corresponding to a complete set of quantum numbers are  Type $M$ operators. Our aim is to determine the values of $M$ for different sectors and to see if these operators conform to the ansatz of Sect.~\ref{Type M}.

As an example, we consider 1d Hubbard model on 6 sites, 3 spin up and 3 spin down electrons. Periodic boundary conditions are assumed. The Hamiltonian reads
\begin{equation}
\hat H=T\sum_{j=1}^6\sum_{s=\up\dn}
(c^\dagger_{js}c_{j+1\,s}+c^\dagger_{j+1\,s}c_{js})+ 
U\sum_{j=1}^6\left(\hat n_{j\uparrow}-
\frac{1}{2}\right)\left(\hat n_{j\downarrow}-
\frac{1}{2}\right),
\label{hub}
\end{equation}
where $c^\dagger_{js}$ and  $c_{js}$ are fermionic creation and 
annihilation operators, respectively, $s$ denotes the spin projection,
$\hat n_{js}=c^\dagger_{js}c_{js}$ is the number operator and $T$ and $U$ are real parameters. There is a hierarchy of conserved currents commuting with each other and with the Hubbard Hamiltonian. The first nontrivial current is linear in both $T$ and $U$, while higher currents are polynomials in $U$ of order 3 and higher.
Following Ref.~\onlinecite{heilmann}, we choose energy units so that $U-4T=1$, which is equivalent to the replacement
$$
U=u\quad T=(u-1)/4, 
$$
where $u$ is a dimensionless real parameter. 

Hamiltonian \re{hub}  for 3 spin up and 3 spin down electrons can be  represented by a square matrix of size $\left(C^6_3\right)^2=400$. It has numerous parameter independent symmetries such as  total momentum, spin, particle-hole symmetry etc. Using these symmetries, one can bring the Hamiltonian to a block diagonal form with block sizes ranging from 1 to 16; see Refs.~\onlinecite{heilmann,emil} for details as well as the quantum numbers corresponding to block sizes  shown in Table~\ref{typH}. With the help of this procedure and algorithms described in Sect.~\ref{numerics} and Appendix~\ref{inverse}, we determine the smallest possible blocks, their type and whether they conform to our ansatz of Sect.~\ref{Type M}. The results are displayed in 
Table~\ref{typH}. The total momentum takes integer values $P=0,\dots,5$ in our notation. Spectra for momenta $P$ and $6-P$ are identical due to a spatial reflection symmetry. For simplicity, we show the results only for momenta $P=1,5$ and $P=2,4$.

\begin{center} 
\begin{table}[hptb]
\begin{tabular}{m{9cm} m{9cm}}

\begin{tabular}{|c|c|c|c|}
\hline
\multicolumn{4}{|c|}{Momenta $P=1,5$} \\
\hline
\# of identical blocks & $N \times N$ & Type & Is it ansatz? \\ \hline
\multirow{1}{*}{1} & $8 \times 8$ & Type 3 & Yes \\
\multirow{1}{*}{2} & $3 \times 3$ & Type 1 & Yes \\
\multirow{1}{*}{2} & $16 \times 16$ & Type 12 & No \\
\multirow{1}{*}{1} & $14 \times 14$ & Type 3 & Yes \\
\multirow{1}{*}{2} & $3 \times 3$ & Type 1 & Yes \\
 \hline
 \end{tabular} & 
 
\begin{tabular}{|c|c|c|c|}
\hline
\multicolumn{4}{|c|}{Momenta $P=2,4$} \\
\hline
\# of identical blocks & $N \times N$ & Type & Is it ansatz? \\ \hline
\multirow{1}{*}{2} & $12 \times 12$ & Type 7 & No \\
\multirow{1}{*}{1} & $14 \times 14$ & Type 11 & No \\
\multirow{1}{*}{2} & $4 \times 4$ & Type 1 & Yes \\
\multirow{1}{*}{2} & $2 \times 2$ & --- & --- \\
\multirow{1}{*}{1} & $16 \times 16$ & Type 6 & No \\
 \hline
 \end{tabular}\\
    \end{tabular} 
 \caption{ \label{typH} Blocks of 1d Hubbard model are matrices of various types, some of which are described by the ansatz of Sect.~\ref{Type M}. Each block corresponds to a complete set of parameter-independent symmetry quantum numbers. The number of nontrivial commuting partners is $N-M-1$, where $M$ is the type. For blocks in the tables this number ranges from 1 to 10, where 10 corresponds to the $14\times 14$ Type 3 block.}
\end{table}
\end{center}
  We make the following observations:
\begin{enumerate}

\item Recall that $N\times N$ Type $M$ matrix has $N-M-1$ nontrivial commuting partners linear in $u$. We note that blocks in Table~\ref{typH} have 1 to 10 such partners. This implies that the 1d Hubbard model has at least 10 conserved currents linear in $u$, while only one such current has been previously identified. All 10 currents can be written in terms of   fermionic creation/annihilation operators $c^\dagger_{js}, c_{js}$ with the help of projectors onto the corresponding sectors. The resulting expressions however might turn out to be nonlocal and rather cumbersome. 

\item The exact solution of Sect.~\ref{Exact} applies to all ansatz blocks in Table~\ref{typH}. Therefore, in all these sectors Bethe's Ansatz solution for the 1d Hubbard model\cite{essler, lieb} -- a large number of coupled nonlinear equations -- reduces to single equation \re{constraint}. It is interesting to see explicitly how such a simplification becomes possible.

\end{enumerate}
In Appendix~\ref{block} we explicitly write down one of the ansatz Hubbard blocks -- the $8 \times 8$ Type 3 block with momentum $P=1$. 

\section{\label{discussion}Discussion}

In this paper, we have defined quantum integrable systems, linear in a parameter $u$, to be operators for which there exist a number of mutually commuting operators similarly linear in $u$. We introduced a classification of these quantum integrable systems according to type, where we defined a Type $M$ operator as belonging to a commuting family with $K=N-M+1$ linearly independent members. For Types 1 and 2, we were able to resolve the corresponding commutation relations, \eref{comm2}, yielding a complete parameterization. For higher types,  we  extended the Type $2$ parameterization  to create an ansatz, see Eqs.~(\ref{Hmel2})-(\ref{dmM}), that parameterizes a subset of all Type $M$ commuting matrices. Moreover these ansatz matrices are exactly solvable through a {\it single} algebraic expression, \eref{constraint}.

In addition to this analytic approach to resolving commutation relations, we also brute forced random numerical solutions. We observed that the number of parameters necessary to specify one of these randomly generated commuting families is $2N+2M-5$, whereas the ansatz involves $2N+M-2$, i.e.
\[
M-3=\left( \mbox{$\#$ of general Type $M$ parameters} \right)-\left( \mbox{$\#$ of ansatz Type $M$ parameters} \right)
\]
When we attempt to find ansatz parameters that can reproduce a randomly generated commuting family's matrix elements, we find that this is possible for Types 1, 2, and 3 only; Types $M>3$ do not appear to be ansatz matrices, in general. Interestingly, when attempting to find ansatz parameters for the blocks of the 1d Hubbard model Hamiltonian, we find that a number of them are indeed ansatz.
Additionally, we looked at the unusual frequent violations of the non-crossing rule in the spectra of Type $M$ matrices and found that the number of (possibly complex) crossings of a Type $M$ matrix is precisely the same as the order of a polynomial in $u$ whose square is a factor of the discriminant of that matrix' characteristic polynomial. The other factor of that discriminant is polynomial common to all members of the commuting family, and its roots correspond to complex values of $u$ for which all $H(u)$ cease to have a full eigen-decomposition, whereby these matrices have nontrivial Jordan canonical form.

The ansatz constitutes a significant step toward a comprehensive theory of parameter dependent finite dimensional integrable models. However, there remains much to be understood about general Type $M$ operators. For example, it is believed that there is some to-be-determined equivalence between  the existence of parameter dependent conservation laws, and the existence of an exact solution. This relationship is on full display with  ansatz Type $M$ matrices. However we could, for the sake of curiosity, consider a Type $N-1$ ``commuting family'' by following the ansatz exactly as written, but taking $M=N-1$. Note that this a very degenerate  family because the number of independent elements is $K=2$, i.e. a single nontrivial operator and the identity. At first glance, such matrices do not seem to be different from most parameter dependent matrices in that their only commuting partner is the identity.   Nevertheless, these Type $N-1$ matrices are exactly solved by  same equations \re{eigened} and 
\re{constraint}. This suggests that conservation laws linear in the parameter are not necessary for an exact solution, though they may be sufficient. 
Interestingly, we find numerically that all ansatz Type $N-1$ matrices have  $N$ commuting currents \textit{quadratic} in 
$u$.
The exact relationship between these two properties remains elusive.

Most interesting is the question of how the ansatz fails to completely parameterize all Type $M$ commuting matrices; i.e. why ansatz Type $M$ commuting families involve $M-3$ fewer parameters than generic Type $M$. Parameter counting alone tells us that the ansatz parameterizes all Type 3 matrices, which we have confirmed numerically. Nevertheless, nothing in the derivation of the ansatz necessitates this Type 3 parametric completeness: this result is a complete (albeit pleasant) surprise. 
But why this surprise coverage of Type 3?   Why does the ansatz completely parameterize Type 2 with a redundancy that can be made manifest through elliptic functions?

Let us start with the last question first and let it be the foundation of the following speculation: Type $M$ commuting matrices \emph{live} on compact Riemann surfaces of genus $g=M-1$. From this point of view, Type 2 matrices live on tori which, when equipped with a complex structure, become Riemann surfaces of genus 1. Working backward, Type 1 lives on the Riemann sphere. Similarly Type 3 lives on a Riemann surface of genus 2, a complex 2-manifold with two handles, Type 4 a genus 3 surface with three handles, etc.

What do we get for this idle speculation? First, the redundancy of Type 2 ansatz parameters detailed in Sect.~\ref{2gauge} involves elliptic functions. One way to view such functions is as doubly periodic functions over the complex plane, but another way to understand them is as the basic functions that live on a fundamental parallelogram, where identification of the parallelogram edges implies that these functions live on a complex torus. This alone is a middling justification for such speculation in that it is limited to Type 2. %We can get more mileage out of this line of speculation, though. 

The Riemann surface conjecture becomes broadly instructive when we further conjecture that ansatz Type $M$ matrices live on the \emph{hyperelliptic} subset of Riemann surfaces of genus $g=M-1$. It is a well known result of the study of Riemann surfaces that general surfaces of genus $g$ have complex structures parameterized by $3g-3$ complex parameters, whereas hyperelliptic Riemann surfaces are much simpler and have complex structures parameterized by $2g-1$ complex parameters \cite{miranda}. If we speculate that the ansatz consistently maps Riemann surfaces of genus $g=M-1$ to Type $M$ commuting families, but that the ansatz is presently restricted to that subset of surfaces that are hyperelliptic, parameter counting tells us that we should be short $g-2$ complex parameters, i.e. that same $M-3$ parameter gap between randomly generated versus ansatz derived Type $M$ commuting families. Moreover, it is well known that most Riemann surfaces are not hyperelliptic, but that all surfaces of genus 0, 1, and 2 are---much the same way that the ansatz does not cover most types, but is complete for Types 1, 2 and (apparently) 3. 

In future work\cite{owusu?} we will show that the eigenvalues of the ansatz Type $M$ matrices belong to the topologically restricted vector space of meromorphic functions on hyperelliptic Riemann surfaces of genus $M-1$. Moreover, the components of the eigenvectors will also be given by meromorphic functions on these surfaces and we will attempt to generalize the manner in which these functions are determined to arbitrary, non-hyperelliptic Riemann surfaces. There we will see that the well-known Riemann-Roch theorem \cite{miranda} justifies the correspondence between a Riemann surface's genus and the commuting family's type. In particular this theorem will show that the more general genus-type relationship is given by $g \geq M-1$, where the typical situation satisfies the lower bound. From this point of view the fact that sometimes $\widetilde{M} \neq M$ (see Sect.~\ref{xings}) is reflected in the fact that $g=\widetilde{M}-1$.

\section{Acknowledgements}

This research was financially supported in part by the National Science Foundation award NSF-DMR-0547769 and by the David and Lucille Packard Foundation.

\appendix

\section{\label{typeMproof}Does the ansatz really parameterize Type $M$?}

Here we prove that the ansatz  of Sect.~\ref{Type M}  indeed yields Type $M$ families of commuting operators as defined in Sect.~\ref{strategy}. Specifically, we need to show that: 1) $H^i(u)$ defined by \eref{him} admit no $u$-independent symmetry, 2)  $[H^i(u), H^j(u)]=0$ for all $i,j\le N-M$ and 3) the linear space ${\cal V}_K$ formed by $H^i(u)$ has dimension $K=N-M+1$,  i.e. the \emph{only} matrices that commute with all $H^i(u)$ are linear combinations thereof.

{\bf 1. Absence of $u$-independent symmetry.} Let us assume such a symmetry $\Omega \neq aI$ exists, i.e. $ [\Omega,H^i(u)]=0$ for all $u$ and $i$, then
\[
[\Omega,V^i]=[\Omega,T^i]=0.
\]
However, for ansatz Type $M$ matrices we can choose a basis \re{vibasis} on ${\cal V}_K$ such that $d^i_k=1/(\lambda_i-\varepsilon_k)$ where, by premise, all $\varepsilon_k$ are distinct. Consequently an ansatz Type $M$ commuting family has a basis in which all diagonal matrices $V^i$ are non-degenerate. It follows then that the only $\Omega$ that commutes with all these $V^i$ is itself a diagonal matrix. By directly computing the commutator we see that for $\Omega$ to commute with all $T^i$
\[
\gamma_{m}\gamma_{n} (\omega_m - \omega_n) \left(\frac{d^i_{m} -d^i_{n} }{\varepsilon_{m}-\varepsilon_{n}}\right) \dfrac{\Gamma_m +\Gamma_n}{2}=0,
\]
where $\omega_m \equiv \Omega_{mm}$ which is only generally satisfied if
\[
\omega_m=\omega_n
\]
for all $m,n=\{1,\dots,N\}$, i.e. if $\Omega$ is a multiple of the identity.

%{\bf Proof of $[H^i(u), H^j(u)]=0$.} 
{\bf 2. Mutual commutativity.} Using the relation
\[
\left(\Gamma_m^2-\Gamma_k^2\right) \left(\Gamma_l^2-1\right)=\left( \varepsilon_m-\varepsilon_k \right) \sum_{s}{\sum_{t}{\dfrac{P_s P_t}{(\lambda_s-\varepsilon_m)(\lambda_s-\varepsilon_k)(\lambda_t-\varepsilon_l)}}},
\] 
and a bit of algebraic manipulation we find that
\begin{multline}
\left[H^i(u),H^j(u) \right]_{kl}=\left[ [V^i,S], [V^j,S] \right]+\left[[V^i,S],W^j \right]-\left[[V^j,S],W^i \right]_{kl}=\\
\\
-\sum_{m}{\sum_{t>s}^{M} \sum_{s=1}^{M} \dfrac{C_{i j k l} \gamma^2_m P_s P_t (\lambda_s-\lambda_t)^2 (\varepsilon_k-\varepsilon_l) (\varepsilon_l-\varepsilon_m) (\varepsilon_k-\varepsilon_m)}{{(\lambda_i-\varepsilon_m)(\lambda_j-\varepsilon_m)(\lambda_s-\varepsilon_k)(\lambda_t-\varepsilon_l)(\lambda_s-\varepsilon_l)(\lambda_s-\varepsilon_m)(\lambda_t-\varepsilon_k)(\lambda_t-\varepsilon_m)}}},
\label{simp2}
\end{multline}
where
\[
C_{i j k l}=\dfrac{\gamma_k \gamma_l (\lambda_i-\lambda_j) }{4(\lambda_i-\varepsilon_k)(\lambda_j-\varepsilon_k)(\lambda_i-\varepsilon_l)(\lambda_j-\varepsilon_l) \left(\Gamma_k+1\right) \left(\Gamma_l+1\right)}. 
\]
Recall that the $\lambda_i$ satisfy \eref{Gaudin1}. Given this, it can be shown, using partial fraction decomposition, that 
\beg
\sum_m{\dfrac{\gamma_m^2}{(\lambda_{i_1}-\varepsilon_m)(\lambda_{i_2}-\varepsilon_m)\dots(\lambda_{i_k}-\varepsilon_m)}}=0,
\label{ided}
\en
provided   $i_1, i_2, \dots, i_k$ are distinct. All terms summed over the index $m$ in \eref{simp2} contain a factor of the form \re{ided} for $k=2, 3$ and 4, where $i_1, i_2, \dots, i_k$  are by construction distinct. Thus  $[ H^i(u), H^j (u)]=0$, for all $i$ and $j$ ranging from 0 to $K-1$.

{\bf 3. Dimension of the vector space formed by $H^i(u)$.} Next, we show that there are no members of this commuting family that are linearly independent of  $H^i(u)$. Toward this end, let us consider a general matrix $H(u)=T+uV$ and specifically the diagonal matrix $V$ and its nonzero matrix elements $d_m$. First, we note that if there exists an $H(u)$ in the commuting family that is linearly independent of the $H^i(u)=T^i+uV^i$, $i=0,\dots,K-1$, but whose $V$ is linearly \emph{dependent} on the $V^i$, then there exists a linear combination of $H^i(u)$ and $H(u)$ that is both $u$-independent and in the commuting family -- a situation precluded by the absence of $u$-independent symmetry demonstrated above.  

More generally, in Ref.~\onlinecite{shastry}, Shastry shows that for arbitrary (not necessarily ansatz) commuting matrices   -- independent of type -- there exists a necessary (but not sufficient) set of constraints on $d_m$ such that $H(u)$ commute with all elements of the commuting family. The constraint is
\beg
d_{j} \, \mu_{j;kl} +  d_{k} \, \mu_{k;lj}+d_{l} \, \mu_{l;jk} +\sum_{i \neq j,k,l}{d_{i} \, \nu_{ijkl}} =0,
\label{mu's and nu's inner product}
\en
where $\nu_{ijkl}$ and $\mu_{j;kl}$ are quantities characteristic of the commuting family as a whole and not specific to any particular one of its members. $\nu_{ijkl}$ is defined as in \eref{nu1} and is $\mu_{j;kl}$ defined as follows:
\[
\mu_{j;kl} \equiv \dfrac{f_{j}^i-f^i_{k}}{d^i_{j}-d^i_{k}}-\dfrac{f_{j}^i-f^i_{l}}{d^i_{j}-d^i_{l}}+S_{jk}^2 S_{jl}^2
-\sum_{m \neq j,k}{S_{jm}S_{mk}S_{kl}S_{lj} \dfrac{d^i_{m}-d^i_{k}}{d^i_{j}-d^i_{k}}}-\sum_{m \neq j,l}{S_{jk}S_{kl}S_{lm}S_{mj} \dfrac{d^i_{m}-d^i_{l}}{d^i_{j}-d^i_{l}}}.
\]
We have used the matrix elements of $H^i(u)$ to define $\mu_{j;kl}$ above, but it turns out that $\mu_{j;kl}$ is independent of index $i$. Let us now consider these general quantities as determined by matrix elements given by Eqs.~(\ref{Hmel2}) and (\ref{gams}). By direct computation it can be shown that, given the Sec.~\ref{Type M} ansatz for Type $M$ matrices
\beg
\begin{array}{l}
\dis \mu_{j;kl}=-\sum_{t>s}^M{\sum_{s=1}^M{\left[ \frac{P_s P_t (\lambda_s-\lambda_t)^2}{16}\left( \frac{\gamma_j^2}{(\lambda_s-\varepsilon_j)(\lambda_t-\varepsilon_j)} \right)^2  \prod_{q=\{k,l \}}{\frac{\gamma_q^2}{(\lambda_s-\varepsilon_q)(\lambda_t-\varepsilon_q)}} \right],}}\\
\\
\dis \nu_{ijkl}=-\sum_{t>s}^M{\sum_{s=1}^M{\left[ \left( \frac{P_s P_t (\lambda_s-\lambda_t)^2}{16}\right)  \prod_{q=\{ i,j,k,l \}}{\frac{\gamma_q^2}{(\lambda_s-\varepsilon_q)(\lambda_t-\varepsilon_q)}} \right]}}.
\nonumber
\end{array}
\en

Now let us compute \eref{mu's and nu's inner product} expressing $d_m$ as $d_{m}=\sum_{s=1}^{N}{\frac{g_s}{\lambda_s-\varepsilon_m}}$ for some $g_s$. Note this  can always be done because the $N\times N$ Cauchy matrix $A_{sm}=(\lambda_s-\varepsilon_m)^{-1}$ is invertible\cite{cauchy}. With $d_m$ so defined, we directly compute \eref{mu's and nu's inner product} to determine the necessary $g_s$ such that $[H^i(u),H(u)]=0$. We find that
\[
\dfrac{1}{2} \sum_{t>s}^M \sum_{s=1}^M{ \sum_{i=1}^N{\left[ \left( g_s \dfrac{\gamma_i^2}{( \lambda_s-\varepsilon_i)^2}\right) \left( \dfrac{P_s P_t (\lambda_s-\lambda_t) \gamma_j^2 \gamma_k^2 \gamma_l^2}{16(\lambda_s-\varepsilon_j)(\lambda_t-\varepsilon_j)(\lambda_s-\varepsilon_k)(\lambda_t-\varepsilon_k)(\lambda_s-\varepsilon_l)(\lambda_t-\varepsilon_l)}\right) \right]}}=0,
\]
which requires that
\beg
\left| \begin{array}{ccc}
1&1&1\\
\sum_{t}{\dfrac{P_t}{\lambda_t-\varepsilon_j}}&\sum_{t}{\dfrac{P_t}{\lambda_t-\varepsilon_k}}&\sum_{t}{\dfrac{P_t}{\lambda_t-\varepsilon_l}}\\
\sum_s{\dfrac{r_s P_s}{\lambda_s-\varepsilon_j}}&\sum_s{\dfrac{r_s P_s}{\lambda_s-\varepsilon_k}}&\sum_s{\dfrac{r_s P_s}{\lambda_s-\varepsilon_l}}
\end{array}
\right|=0,
\label{det nu}
\en
where summations over $s$ and $t$ run from 1 to $M$ and $r_s=g_s \sum_{i=1}^N{\gamma_i^2/(\lambda_s-\varepsilon_i)^2}$. Note that terms containing $g_s$ with $s>M$ cancel from \eref{mu's and nu's inner product} by virtue of identity \re{ided}.

For  the determinant  to vanish, the rows in the matrix in \eref{det nu} must be linearly dependent. By the uniqueness of partial fractional decomposition it follows that $r_s=r$, i.e. the determinant vanishes if (and only if) $r_s$ does not depend on $s$. Thus we require that 
\beg
g_s=r \left( \sum_{i=1}^N{\dfrac{\gamma_i^2}{(\lambda_s-\varepsilon_i)^2}} \right)^{\lefteqn{-1}}, \quad \mbox{for} \quad 1\le s \leq M.
\label{dd}
\en
It turns out that the degree of freedom associated with $r$ amounts to variation in an overall trace as it can be shown that 
\beg
\widetilde{d}\equiv\sum_{j=1}^{N}{ \dfrac{r}{\lambda_j-\varepsilon_m} \left( \sum_{i=1}^{N}{\frac{\gamma_i^2}{(\lambda_j-\varepsilon_i)^2}} \right)^{ -1}}=rB^{-1},
\en
independent of the index $m$. Thus, it follows that the only $V$ allowed in an ansatz Type $M$ family are constrained to be of the form
\beg
V=\sum_{i=1}^{K}{a_i V^i}+\widetilde{d} I,
\en
where $a_i=g_{i+M}-\widetilde{g}_{i+M}$ and $\widetilde{g}_{i+M}$ is defined by the right hand side of \eref{dd} with $s=i+M$, whereby the ansatz necessarily parameterizes matrices of \emph{exactly} Type $M$. 

\section{\label{inverse}Inverse problem: determining parameters given an ansatz matrix}

Here we detail an algorithm that, given an arbitrary $N\times N$ symmetric matrix $H(u)=T+uV$, determines if it conforms to the ansatz of Sect.~\ref{Type M} and returns the ansatz parameters when it does.

The algorithm is based on the central observation that the difference between a Type 1 $S_{ij}$, see \eref{S1}, and that of an ansatz Type $M$ matrix, see \eref{full form M}, is the factor $\eta_{ij} \equiv (\Gamma_i +\Gamma_j)/2$. If a Type $M$ commuting family has an antisymmetric matrix $S$ for which no such factor can be found, we know it does not conform to an ansatz parameterization. If, however, such a factor exists and it can be determined, then we know that 
\beg
\Sigma_{ij} \equiv \dfrac{S_{ij}}{\eta_{ij}} = \dfrac{\gamma_i \gamma_j}{\varepsilon_i-\varepsilon_j},
\label{sigma}
\en
for some $\gamma_j$ and $\varepsilon_j$ and $\Sigma_{ij}$ so defined satisfies the equation
\beg
\Sigma_{ij} \Sigma_{jk} \Sigma_{kl} \Sigma_{li}+\Sigma_{ik} \Sigma_{kl} \Sigma_{lj} \Sigma_{ji}+\Sigma_{il} \Sigma_{lj} \Sigma_{jk} \Sigma_{ki}=0.
\label{nuvo}
\en
Moreover the reverse is true, i.e. it can be shown that \eref{nuvo} implies \eref{sigma} for some $\gamma_j$ and $\varepsilon_j$. Consequently, if \eref{nuvo} can be solved for $\Gamma_j$, we can essentially strip an ansatz $S_{ij}$ of its factor $\eta_{ij}$, and the ansatz parameters $\gamma_j$ and $\varepsilon_j$ can be determined. Without loss of generality we can multiply \eref{nuvo} by an overall factor of $16 \, \eta_{ij} \eta_{ik} \eta_{il} \eta_{jk} \eta_{jl} \eta_{kl}$ and look for solutions $\Gamma_j$ to resulting equation
\begin{multline}
\left(\Gamma_i+\Gamma_k \right) \left(\Gamma_j+\Gamma_l \right) S_{ij} S_{jk} S_{kl} S_{li}+\left(\Gamma_i+\Gamma_l \right) \left(\Gamma_j+\Gamma_k \right) S_{ik} S_{kl} S_{lj} S_{ji} +
\\
\left(\Gamma_i+\Gamma_j \right) \left(\Gamma_k+\Gamma_l \right) S_{il} S_{lj} S_{jk} S_{ki}=0
\label{xi}
\end{multline}
This is a massively overdetermined set of $N(N-1)(N-2)(N-3)/4!$ equations, quadratic in the $N$ terms $\Gamma_j$. We note that the equations are actually linear and homogenous in the terms $B_{ij} \equiv \Gamma_i \Gamma_j$ and that it is possible to attempt solving for the $\Gamma_j$ by first finding the minimal set of linearly independent linear equations in $B_{ij}$ and then solving for $\Gamma_j$. In numerical practice, however, Mathematica's NSolve -- programmed to find all sets $\Gamma_j$ that satisfy the $N(N-1)(N-2)(N-3)/4!$ equations of \eref{xi} -- quickly finds nontrivial values for the $\Gamma_j$ directly from the overdetermined quadratic equations, if they exist. We find that for matrices with a generic, non-ansatz $S$, solving \eref{xi} yields only trivial solutions, e.g. $\Gamma_j=0$, for all $j$ but one. For ansatz matrices where $M \geq 3$, however, there are two equivalent, inversely related non-trivial solutions to \eref{xi}, i.e.
\[
\Gamma_i= \{\alpha D_i, \beta D_i^{-1} \}
\]
where $\alpha,\beta$ are arbitrary complex numbers. Note that having real $S_{ij}$ does not generally guarantee that the corresponding $\Gamma_i$ are themselves real. Note also that, despite having proved that Type 1 and 2 matrices always have an ansatz parameterization, one can just as well attempt to determine their $\Gamma$'s using solutions to \eref{xi}. However, in both cases the $\Gamma_j$ cannot be determined uniquely. In the Type 1 case, NSolve generates a warning indicating that there are not enough constraints for it to express a solution set and that it must generate three additional constraints in order to proceed; from Sect.~\ref{Type M} we know that these additional constraints amount to fixing the threefold parametric redundancy unique to Type 1. In the case of Type 2, NSolve generates a warning that it must generate a single constraint to proceed; this corresponds to a single redundancy in the parameterization of Type 2 matrices involving elliptic functions, which we detailed in Sect.~\ref{2gauge}.

If nontrivial solutions $\Gamma_j$ exist, the above procedure determines their values.
It follows from \eref{sigma} that the resulting $\Sigma_{ij}$ satisfy a corollary equation 
 \beg
 (\gamma_i \Sigma_{jk})^{-1}-(\gamma_j \Sigma_{ik})^{-1}+(\gamma_k \Sigma_{ij})^{-1}=0,
 \en
for some $\gamma_i$ to be determined. This we do by fixing $i$ and $j$ and choosing arbitrary values for $\gamma_i$ and $\gamma_j$ such that we can use equation
\beg
\gamma_k^{-1}=\Sigma_{jk} \left( \left(\gamma_j \Sigma_{ik} \right)^{-1}- \left( \gamma_i \Sigma_{jk} \right)^{-1} \right).
\en
to determine $\gamma_k$ for all $k \neq i,j$. From here, determining $\varepsilon_k$ is straightforward, i.e.
\beg
\varepsilon_k=\varepsilon_i-\dfrac{\gamma_i \gamma_k}{ \Sigma_{ij}}, \quad \mbox{for all} ,\quad k \neq  i
\en
where we have yet another arbitrary degree of freedom in our choice of $\varepsilon_i$. These three parametric redundancies, unearthed by this algorithm, constitute an interesting gauge freedom in all ansatz Type $M$ commuting families, see Sect.~\ref{Gauge}. 

If and when the algorithm finds values $\gamma_j,\varepsilon_j$ and $\Gamma_j$ for a putative ansatz Type $M$ matrix $H(u)$, proceeding to determine ansatz parameters $\lambda_j$ is a matter of determining whether there exists an $\widetilde{H}(u)=u \widetilde{V} +\widetilde{T}$ in the commuting family wherein the nonzero elements of diagonal matrix $\widetilde{V}$ are of the form 
\[
\widetilde{d}_j=\dfrac{1}{\widetilde{\lambda}-\varepsilon_j}.
\]
for $\widetilde{\lambda}$ to be determined by the algorithm, see Sect.~\ref{Type M}. Recall from Sect.~\ref{numerics} that in determining the size of a random Type $M$ commuting family, finding all matrices that commute with $H(u)$ reduces to $N+M-2$ linearly independent equations linear in its commuting partner's diagonal elements. Of these equations, we know that because there are only $N-M+1$ independent members of the commuting family, exactly $M-1$ of these equations can be found that involve the $N$ $\widetilde{d}_j$ alone. Consequently there will be $M-1$ constraints on $\widetilde{\lambda}$ of the form
\[
\sum_{j=1}^{N}{\dfrac{c^k_j}{\widetilde{\lambda}-\varepsilon_j}}=0,
\]
where coefficients $c^k_j$ are determined by the matrix elements of $H(u)$. Solving these simultaneous equations reduces to finding all $\widetilde{\lambda}$ that satisfy $M-1$ polynomials of order $N-1$. Generally, such simultaneous equations polynomial have no solution. These ones derived from ansatz matrices, however, have an $M$-element solutions set $\{\lambda_1,\lambda_2,\dots \lambda_M \}$. To be ansatz parameters, each one must correspond to the same parameter $B$ satisfying \eref{Gaudin1}.

Finding the rest of the $M$ parameters $P_j$ is a matter of solving $N$ overdetermined linear equations
\[
\Gamma_k^2=1+\sum_{j=1}^{M} \frac{P_j}{\lambda_{j}-\varepsilon_k},
\]
see \eref{gams}. If the algorithm can determine these $P_k$, $H(u)$ is ansatz. Failure to determine any of these parameters uniquely (up to the aforementioned gauge redundancy), or any inconsistency with respect to the ansatz equations indicates that the matrix is not ansatz. For example, it is possible that the algorithm could find some $\Gamma_j$ and parameters $\varepsilon_j,\gamma_j$ given some matrix, and yet fail to have a consistent ansatz parameter $B$. In practice, however, all matrices tested through this algorithm failed to yield ansatz parameters at the $\Gamma$-stage in so much as the algorithm could not find nontrivial $\Gamma_j$ consistent with \eref{xi}. That is, if there are non-ansatz Type $M$ matrices that satisfy \eref{full form} for some $\Gamma_j$ not satisfying \eref{gams}, they appear to be rather rare in a random ensemble of Type $M$ matrices.

\section{\label{block}An example of a sector of the 1d Hubbard model described by the ansatz}

Here we explicitly write down the $8\times 8$, momentum $P=1$  block of the 1d Hubbard model, which is an  ansatz Type 3 matrix, see Sect.~\ref{Taxonomy}. This block is of the form $T+uV$,
where $V$ is diagonal matrix with the following entries:
\beg
\mbox{diagonal of $V$}=\frac{1}{4}\Bigl(-3,  1, -3, 0,  3, 0,   3,   -1\Bigr)
 \en
and
  \beg
T=\frac{1}{12}\left(
\begin{array}{cccccccc}
7 & \sqrt{6} & 0& 1& -1 & 3 & -3& \sqrt{6}\\
\sqrt{6} & -3& \sqrt{6}&\sqrt{6}& -\sqrt{6} & -\sqrt{6} & -\sqrt{6} &0\\
0&\sqrt{6}&11&3&3&1&1&-\sqrt{6}\\
1&\sqrt{6}& 3&-2&-1&0&3&\sqrt{6}\\
-1&-\sqrt{6}&3&-1&-11&3&0&-\sqrt{6}\\
3&-\sqrt{6}&1&0&3&2&-1&\sqrt{6}\\
-3&-\sqrt{6}&1&3&0&-1&-7&\sqrt{6}\\
\sqrt{6}&0&-\sqrt{6}&\sqrt{6}&-\sqrt{6}&\sqrt{6}&\sqrt{6}&3\\
  \end{array}
\right).
\en
Using the procedure outlined in Appendix~\ref{inverse}, we find  the corresponding ansatz parameters
\[
\varepsilon=\{-1, 0, 5, -5/2, 5, -5/2, -1, 0\},
\]
\[
\gamma=\frac{1}{2\sqrt{6}} \left\{1-i,  3\sqrt{6}+i\sqrt{6}, -1-7i, 4+8i, -7+i, -8+4i, 1+i, \sqrt{6}-3\sqrt{6}i\right\},
\]
and
\[
\Gamma_j^2=\dfrac{5\varepsilon_j^3-\left( 1 - 37 i \right) \varepsilon_j+ 4 - 3 i   }{5\varepsilon_j^3-\left(  1 + 37 i \right) \varepsilon_j+ 4 + 3 i  }.
\]


\begin{thebibliography}{99}

  \bibitem{Caux} J.-S. Caux, J. Mossel, J. Stat. Mech.  P02023 (2011).

\bibitem{arnold} V. I. Arnold, \emph{Mathematical Methods of Classical Mechanics}, (Springer-Verlag, New York, 1978).

\bibitem{Baum} H. Baumgrtel, \emph{Analytic perturbation theory for matrices and operators}, p. 89 (Birkhuser Verlag, 1985).

%\bibitem{arg} See e.g. Sect. 6 in Ref.~\onlinecite{emil}.

\bibitem{hubbard} J. Hubbard, Proc. Roy. Soc. A 276 238 (1963).

\bibitem{essler} F. H. L. Essler , H. Frahm, F. G\"ohmann, A. Kl\"umper, and V. Korepin, \textit{The One-Dimensional Hubbard 
Model} (Cambridge University Press, Cambridge, 2005).

\bibitem{owusu} H. K. Owusu, K. Wagh, and E. A. Yuzbashyan,  J. Phys. A: Math. Theor.  42, 035206 (2009).

\bibitem{sklyanin} E. Sklyanin, J. Sov. Math. 47, 2473 (1989); Progr. Theoret. Phys. Suppl. 118, 35 (1995).

\bibitem{bcs} J. Bardeen, L.N. Cooper, and J.R. Schrieffer, Phys. Rev. 108, 1175 (1957).

\bibitem{Gaudin}  M. Gaudin, Note CEA   1559, 1 (1972); J. Phys. (Paris)   37, 1087 (1976);
 \emph{La fonction d'onde de Bethe}, (Masson, Paris, 1983).

 \bibitem{cambiaggio} Cambiaggio, M. C., A. M. F. Rivas, and M. Saraceno,  Nucl. Phys. A 624, 157 (1997).


\bibitem{dukelsky} J. Dukelsky, S. Pittel, G. Sierra, Rev. Mod. Phys. 76, 643 (2004).

\bibitem{S} B. S. Shastry,  Phys. Rev. Lett.  56, 1529 (1986); ibid. 56, 2453 (1986); J. Stat. Phys. 50, 57 (1988).


\bibitem{Lu}

M. L\"uscher,  Nucl. Phys.  B117, 475 (1976).

\bibitem{Gr}

H. Grosse,   Lett. Math. Phys.  18, 151 (1989).

\bibitem{GM}

M. P. Grabowski and P. Mathieu,   Ann. Phys. 243, 299 (1995).

\bibitem{zhou} H. Zhou, L. Jiang, and J. Tang, J. Phys. A: Math. Gen 23, 213 (1990).

\bibitem{Fu} B. Fuchssteiner, in \textit{Symmetries and Nonlinear Phenomena}, p. 22Ð50 (World Scientific Publishers, Singapore, 1988).



\bibitem{count} Up to an arbitrary $u$-independent orthogonal transformation (e.g. to a basis where $T$ is diagonal), which adds $N(N-1)/2$ parameters in both cases. 
 

\bibitem{shastry} B. S. Shastry,  J. Phys. A: Math. Gen.  38, 431 (2005).

\bibitem{emil} E. A. Yuzbashyan, B. L. Altshuler and B. S.
Shastry, J. Phys. A: Math. Gen. 35, 7525 (2002).

\bibitem{shastry1} B. S. Shastry,  J. Phys. A: Math. Theor. 44, 052001 (2011).

\bibitem{Hund} F. Hund, Z. Phys.   40, 742 (1927).

\bibitem{Neumann} J. von Neumann, E. Wigner, Z. Phys.  30,
467 (1929).


\bibitem{Teller} E. Teller, J. Phys. Chem.  41, 109 (1937).

\bibitem{Landau}L. D. Landau and E.M. Lifshitz, \emph{Quantum Mechanics:
Non-Relativisitic Theory}, pp. 304-305 (Pergamon Press, Oxford, 1980).


\bibitem{Longuet-Higgins}H. C. Longuet-Higgins, Proc. R. Soc. A  344, 147 (1975).

\bibitem{Naqvi}K. R. Naqvi, W. B. Brown, Int. J. Quantum Chem.
6, 271 (1972).

\bibitem{Kestner}J. P. Kestner, L.-M. Duan, Phys. Rev.
A 76, 033611 (2007).



\bibitem{normalization} We choose this condition because with the four parameters there is a redundant overall scale. Choosing $qt-rs=1$ fixes that scale and, as it turns out, simplifies a number of calculations. 


\bibitem{type2}  As explained in Sect.~\ref{strategy}, in the canonical basis each Type $M$ basic operator at $u=\infty$ is a diagonal matrix with $N-M$ zero eigenvalues.

\bibitem{cauchy} S. Schechter,  Mathematical Tables and Other Aids to Computation 13 (66), 73 (1959).


\bibitem{Abramowitz} M. Abramowitz and I. A. Stegun,   \textit{Handbook of Mathematical Functions with Formulas, Graphs, and Mathematical Tables},  Ch. 16,  pp. 567-581  (New York: Dover, 1972).


\bibitem{wolf} S. Wolfram, \emph{The Mathematica Book} (Wolfram Media, Champaign, IL, USA, fifth edition, 2003).



\bibitem{heilmann} 
O. J. Heilmann  and E. H. Lieb,   Ann. N. Acad. Sci. 172, 583 (1971).


\bibitem{Horn} R. A. Horn and C. R. Johnson, \emph{Matrix Analysis}, pp. 121-142 (Cambridge University Press, Cambridge, 1985).



\bibitem{lieb} E. H. Lieb and F. Y. Wu,   Phys. Rev. Lett. 20, 1445 (1968); Erratum, ibid. 21, 192 (1968).

\bibitem{owusu?} H. Owusu, unpublished.



 \bibitem{miranda} R Miranda, \emph{Algebraic Curves and Riemann Surfaces} (Providence, RI, American Mathematical Society, 1995).

\end{thebibliography}
\end{document}